\documentclass[%
 reprint,
superscriptaddress,
nofootinbib,
 amsmath,amssymb,
 aps,
 prd,
]{revtex4-2}

\usepackage[usenames,dvipsnames]{xcolor}
\clearpage{}%

\usepackage{graphicx}%
\graphicspath{{./}{figures/}}
\usepackage[caption=false,labelformat=simple]{subfig}

\usepackage{dcolumn}%
\usepackage{booktabs}
\usepackage{bm}%
\usepackage{hyperref}%
\hypersetup{
	colorlinks=true,
	citecolor=NavyBlue,
	linkcolor=NavyBlue,
	urlcolor=NavyBlue,
}
\usepackage{cancel}
\usepackage{slashed}
\usepackage[normalem]{ulem}
\usepackage{braket}
\usepackage{nicefrac}
\usepackage[acronym]{glossaries}
\newacronym{GSN}{GSN}{generalized Sasaki-Nakamura}
\newacronym{SN}{SN}{Sasaki-Nakamura}
\newacronym{MST}{MST}{Mano-Suzuki-Takasugi}
\newacronym{BH}{BH}{black hole}
\newacronym{GW}{GW}{gravitational wave}
\newacronym{ODE}{ODE}{ordinary differential equation}
\newacronym{PDE}{PDE}{partial differential equation}
\newacronym{GDT}{GDT}{Generalized Darboux Transformation}
\newacronym{QNM}{QNM}{quasinormal mode}
\newacronym{QNEF}{QNEF}{quasinormal excitation factor}
\newacronym{EMRI}{EMRI}{extreme mass-ratio inspiral}
\newacronym{LIGO}{LIGO}{Laser Interferometer Gravitational-Wave Observatory}
\newacronym{LISA}{LISA}{Laser Interferometer Space Antenna}
\newacronym{DECIGO}{DECIGO}{Deci-hertz Interferometer Gravitational wave Observatory}
\newacronym{AD}{AD}{automatic differentiation}
\newacronym{CSV}{CSV}{comma-separated values}

\makeatletter
\newcommand*{\glsplainhyperlink}[2]{%
  \colorlet{currenttext}{.}%
  \colorlet{currentlink}{\@linkcolor}%
  \hypersetup{linkcolor=currenttext}%
  \hyperlink{#1}{#2}%
  \hypersetup{linkcolor=currentlink}%
}
\let\@glslink\glsplainhyperlink
\makeatother

\newcommand{\etal}{{\it et al.}}

\newcommand{\plotinstr}{Modes with $m \geq 0$ are plotted with solid lines, while $m < 0$ modes are plotted with dashed lines. In addition, modes with the same value of $|m|$ are plotted with the same color and connected at $a/M = 0$, and the value of $a/M$ increases away from that connection point. Another set of \glspl{QNM} can be obtained by simply ``mirroring'' the modes shown here along the imaginary $M\omega$ axis and thus not plotted explicitly. Note that this plot is intended to look for general trends and anomalies in those \gls{QNM} trajectories and inspect our calculations for numerical artifacts. Numerical values of these \gls{QNM} frequencies are available online \cite{qnmqnefdata}.}

\begin{document}

\title{Quasinormal modes and excitation factors of Kerr black holes}

\author{Rico K.~L. Lo}
\email{kalok.lo@nbi.ku.dk}
\author{Leart Sabani}
\email{leart.sabani@nbi.ku.dk}
\affiliation{%
Center of Gravity, Niels Bohr Institute, Blegdamsvej 17, 2100 Copenhagen, Denmark}%
\author{Vitor Cardoso}
\affiliation{%
Center of Gravity, Niels Bohr Institute, Blegdamsvej 17, 2100 Copenhagen, Denmark}
\affiliation{%
CENTRA, Departamento de F\'{\i}sica, Instituto Superior T\'ecnico---IST, Universidade de Lisboa---UL, Avenida Rovisco Pais 1, 1049-001 Lisboa, Portugal}

\date{\today}%

\begin{abstract}
Theoretical understanding of the characteristic oscillations of a perturbed black hole, also referred to as quasinormal modes (QNMs), is crucial to interpreting the late stage of binary black hole mergers that we now routinely observe in gravitational wave detectors.
In this work, we introduce a new approach, based on the generalized Sasaki-Nakamura formalism, to compute the QNM spectra and their excitation factors (QNEFs), for scalar, electromagnetic, and gravitational perturbations.
Using this approach, QNM wave functions remain finite at the horizon and spatial infinity.
Our results, in general, agree with previous calculations that were performed using different methods, though we further clarify that QNEFs and their scaling with the mass of the black hole depend on the spin weight of the perturbation.
We show that avoided crossing is {\it not} a general phenomenon: The real or the imaginary part of the eigenvalues can cross each other but not both simultaneously, and, when crossing occurs in one part, repulsion follows in another part. Eigenvalue repulsion, originating from branch point singularities, still plays an important role in the QNM spectra, despite the fact that the spectra depend only on the real-valued black hole angular momentum.
\end{abstract}

\maketitle

\section{Introduction \label{sec:intro}}
The geometry of an uncharged, rotating \gls{BH} in vacuum is described by the Kerr metric~\cite{Kerr:1963ud} and is characterized by its mass $M$ and the angular momentum per unit mass $a$ of the \gls{BH}.
When a \gls{BH} is disturbed, it emits \glspl{GW} during the relaxation process to a new equilibrium configuration. At late times, the dynamics are well described by a linearization of Einstein equations. The characteristic frequencies of the corresponding problem are called the \glspl{QNM} of a \gls{BH}; they are the characteristic oscillations excited whenever the \gls{BH} interacts with its exterior. For instance, the excited remnant formed after the merger of two \glspl{BH} emits gravitational radiation composed primarily of \glspl{QNM} to relax back to its stationary state. This was first detected by the \acrlong{LIGO} in 2015 \cite{LIGOScientific:2016aoc}.
With advancements in the sensitivity of these detectors, we might soon be able to perform \gls{BH} spectroscopy~\cite{Berti:2005ys,Berti:2016lat} measuring a number of distinct characteristic frequencies in the spectrum from \gls{GW} data. The tantalizing prospect of actually doing \gls{BH} spectroscopy, and the potential to uncover new physics, deserves a careful look at the problem, and indeed there is vibrant activity in this topic~\cite{Berti:2009kk,Baibhav:2023clw}.

\subsection{Primer on black hole perturbation theory}
Because of the lack of spherical symmetry of Kerr \glspl{BH}, one would have speculated that the partial differential equations governing the linear perturbation of a field would not be separable; i.e., the equation would be, in general, a \acrlong{PDE} of some variables, say, the Boyer-Lindquist $(t,r,\theta,\phi)$ coordinates \cite{Boyer:1966qh, Misner:1973prb}. However, by considering instead some projections of the field, separation of variables on the governing wave equation of those scalar quantities can, surprisingly, be achieved~\cite{Teukolsky:1973ha}. 
The scalar quantity $\psi^{(s)}$ to use depends on the nature of the perturbed field, where the bracketed superscript $s$ denotes its spin weight of the field ($s=0,\pm1,\pm2$ for scalar fields, electromagnetic fields, and gravitational fields, respectively). 
If we further assume a harmonic time dependence $\sim e^{-i \omega t}$ (where $\omega$ for now is just a variable), then a wave function $\psi^{(s)}$ can be written schematically as 
\begin{equation}
\label{eq:psi}
	\psi^{(s)}(t, r, \theta, \phi) = \sum_{\omega} \sum_{\ell m} {}_{s}R_{\ell m \omega}(r) {}_{s}S_{\ell m \omega}(\theta, \phi) e^{-i\omega t},
\end{equation}
where the indices $\ell, m$ label angular wave functions ${}_{s}S_{\ell m\omega}(\theta, \phi)$ and ${}_{s}R_{\ell m \omega}(r)$ denotes a (time-independent) radial wave function. Focusing on the radial sector, the \gls{ODE} that a radial wave function satisfies can be written as
\begin{equation}
\label{eq:schematic_radial_Teukolsky_eq}
    \mathcal{L} \left[ {}_{s}R_{\ell m \omega} \right] = \mathcal{S},
\end{equation}
where $\mathcal{L}$ is a second-order linear differential operator representing the radial Teukolsky equation \cite{Teukolsky:1973ha} and $\mathcal{S}$ denotes the source term driving the perturbations. 
A number of techniques to solve the above equation have been developed over the years.
One of those techniques is the \gls{SN} formalism \cite{SASAKI198268, Hughes:2000pf, Lo:2023fvv}, where Eq.\eqref{eq:schematic_radial_Teukolsky_eq} is transformed into a new \gls{ODE} that is more suitable for direct numerical integrations.
If ${}_{s}X_{\ell m \omega}$ denotes a solution to this new \gls{ODE} and ${}_{s}\Lambda$ denotes an operator that transforms a radial Teukolsky solution $R$ to the corresponding $X$ solution (suppressing the $s \ell m \omega$ subscripts whenever possible therein), then Eq.~\eqref{eq:schematic_radial_Teukolsky_eq} can be recast simply as
\begin{equation}
\label{eq:schematic_radial_Teukolsky_eq_with_SN}
    \mathcal{L} \left[ {}_{s}\Lambda^{-1} \left[ {}_{s}X_{\ell m \omega} \right] \right] = \mathcal{S},
\end{equation}
where ${}_{s}\Lambda^{-1}$ denotes the inverse operator of ${}_{s}\Lambda$ \cite{GWPT, Lo:2023fvv}.

To solve for generic cases where $\mathcal{S} \neq 0$, one can deploy the Green's function method \cite{ARFKEN2013447}. Suppose $u$ represents either ${}_{s}R_{\ell m \omega}$ or ${}_{s}X_{\ell m \omega}$, $u_{\text{in,out}}^{\infty}$ and $u_{\text{in,out}}^{\text{H}}$ denote solutions that solve Eq.~\eqref{eq:schematic_radial_Teukolsky_eq} or \eqref{eq:schematic_radial_Teukolsky_eq_with_SN} with $\mathcal{S}=0$ and satisfy the purely in- or outgoing boundary condition at spatial infinity and at the horizon, respectively. Furthermore, we construct two linearly independent solutions $u_{\text{in}}$ and $u_{\text{up}}$ from these solutions as
\begin{equation}
\label{eq:u_in}
    u_{\text{in}} = \begin{cases}
    u_{\text{in}}^{\text{H}} & \text{as } r_* \to -\infty, \\
    B^{\text{inc}} u_{\text{in}}^{\infty} + B^{\text{ref}} u_{\text{out}}^{\infty} & \text{as } r_* \to +\infty,
    \end{cases}
\end{equation}
where the notation of $A_{\text{in}} \equiv B^{\text{inc}}$ and $A_{\text{out}} \equiv B^{\text{ref}}$ is also being used in literature on \glspl{QNM}\footnote{We will use both notations interchangeably but never mix the two notations (so either $A_{\text{in, out}}$ or $B^{\text{inc, ref}}$ will appear together).} and
\begin{equation}
\label{eq:u_up}
    u_{\text{up}} = \begin{cases}
    C^{\text{ref}} u_{\text{in}}^{\text{H}} + C^{\text{inc}} u_{\text{out}}^{\text{H}} & \text{as } r_* \to -\infty, \\
    u_{\text{out}}^{\infty} & \text{as } r_* \to +\infty.
    \end{cases}
\end{equation}
Here $r_{*}$ is the tortoise coordinate related to the Boyer-Lindquist $r$ coordinate by
\begin{equation}
\label{eq:rstar_from_r}
    r_{*}(r) = r + \frac{2Mr_{+}}{r_{+}-r_{-}} \ln \left( \frac{r - r_+}{2M} \right) - \frac{2Mr_{-}}{r_{+}-r_{-}} \ln \left( \frac{r - r_-}{2M} \right),
\end{equation}
where $r_{\pm} = M \pm \sqrt{M^2 - a^2}$ is the outer and the inner horizon of the \gls{BH}, respectively. More importantly, this coordinate transformation maps $r = r_{+}$ into $r_{*} \to -\infty$, while $r \to \infty$ maps into $r_{*} \to \infty$.

More explicitly, the solution ${}_{s}R_{\ell m \omega}$ of Eq.~\eqref{eq:schematic_radial_Teukolsky_eq} with $\mathcal{S} \neq 0$ on the right-hand side that satisfies the purely ingoing boundary condition at the horizon and purely outgoing boundary condition at spatial infinity can then be written, according to the theory of Green's functions \cite{ARFKEN2013447}, as
\begin{multline}
\label{eq:inhomogeneous_sol}
{}_{s}R_{\ell m \omega}(r) =  \dfrac{R_{\text{in}}(r)}{\mathcal{W}_{R}} \int_{r}^{\infty} dr'\,\dfrac{R_{\text{up}}(r') \mathcal{S}(r')}{\Delta^{-s}(r')} \\
	+ \dfrac{R_{\text{up}}(r)}{\mathcal{W}_{R}} \int_{r_{+}}^{r}dr'\,\dfrac{R_{\text{in}}(r') \mathcal{S}(r')}{\Delta^{-s}(r')},
\end{multline}
where we have made the substitution of $u \mapsto R$ and $\Delta(r) \equiv r^2 - 2Mr + a^2$.
In the limit $r \to \infty$ (appropriate for observers far away from the perturbed \gls{BH}, e.g., us), using the expression of $\mathcal{W}_{R}$ and $R_{\text{up}}(r)$ in Ref.~\cite{Lo:2023fvv}, we have
\begin{multline}
\label{eq:farfield_R}
	R(r \to \infty) = \\
	\left[ \dfrac{A_{\mathrm{out}}^{\mathrm{T}}}{2i\omega A_{\mathrm{in}}^{\mathrm{T}}}  \int_{r_{+}}^{\infty}dr'\,\dfrac{R_{\text{in}}(r') \mathcal{S}(r')}{A_{\mathrm{out}}^{\mathrm{T}} \Delta^{-s}(r')}\right] r^{-(2s+1)}e^{i\omega r_{*}},
\end{multline}
where a superscript T is used to denote quantities related to Teukolsky solutions (similarly, we use a superscript SN for quantities related to \gls{SN} solutions). Notice that the term enclosed in the square bracket is independent of the field evaluation point $r$.

Putting Eq.~\eqref{eq:farfield_R} into Eq.~\eqref{eq:psi} to get the full wave function $\psi^{(s)}(t,r \to \infty,\theta,\phi)$ in the far field limit, we have
\begin{multline}
	\psi^{(s)}(t, r \to \infty,\theta,\phi) = \sum_{\ell m} \int_{-\infty}^{+\infty} d\omega\, e^{i\omega r_{*} - i\omega t} {}_{s}S_{\ell m \omega}(\theta, \phi) \\ \left[ \dfrac{A_{\mathrm{out}}^{\mathrm{T}}}{2i\omega A_{\mathrm{in}}^{\mathrm{T}}}  \int_{r_{+}}^{\infty}dr'\,\dfrac{R_{\text{in}}(r') \mathcal{S}(r')}{A_{\mathrm{out}}^{\mathrm{T}} \Delta^{-s}(r')}\right] r^{-(2s+1)}.
\end{multline}
In particular, we replace the schematic summation over $\omega$ in Eq.~\eqref{eq:psi} with the actual integral over $\omega$.
Instead of evaluating the integral along the real line, one can evaluate an equivalent contour integral on the complex-$\omega$ plane \cite{ARFKEN2013469}.\footnote{Moreover, we choose a semicircular contour going over the lower half of the complex-$\omega$ plane with a branch cut starting at the origin along the negative imaginary axis. For an illustration, see Fig.~2 in Ref.~\cite{Berti:2006wq}.}
Thanks to the residue theorem, part of the contour integral reduces to a much simpler discrete sum over the residues evaluated at the poles of the analytic integrand \cite{Leaver:1986gd}.
At some particular values of $\omega$ referred to as \gls{QNM} frequencies $\omega_{\text{QNM}}$, $A_{\text{in}}^{\text{T}} \equiv B^{\text{inc}}_{\text{T}}$ that appears in the denominator of the integrand vanishes (in fact, so as $C^{\text{inc}}_{\text{T}}$). Mathematically, these are the poles of the Green's functions and will contribute to the sum. Physically, at those frequencies, the wave functions become purely ingoing at the horizon and outgoing at spatial infinity.

Let us write $A_{\text{in}}(\omega)$ near those \gls{QNM} frequencies as
\begin{equation}
	A_{\text{in}}\left(\omega\right) = \cancelto{0}{A_{\text{in}}\left(\omega_{\text{QNM}} \right)} + \underbrace{\dfrac{dA_{\text{in}}}{d\omega}}_{\neq 0} \left( \omega - \omega_{\text{QNM}} \right) + \dots,
\end{equation}
and we can then schematically write the sum over the \gls{QNM} frequencies as
\begin{equation}
\label{eq:QNM_contribution}
	\psi^{(s)} \supset \sum_{\omega_n}	2\pi i \dfrac{A_{\mathrm{out}}(\omega_n)}{2i\omega_n dA_{\mathrm{in}}(\omega_n)/d\omega} E_{n},
\end{equation}
where $n = 0, 1, \dots$ is the overtone number that labels the \gls{QNM} frequencies by ascending $|\text{Im}\;\omega_n|$ and $E_n$ depends on the source term. For the particular $s \ell m$ mode under consideration, $\omega_0$ (or $n = 0$) is referred to as the fundamental mode and $\omega_{1,2,\dots}$ (or $n=1,2,\dots$) are referred to as the overtones.
Following the notations in Ref.~\cite{Berti:2006wq}, we define the part in Eq.~\eqref{eq:QNM_contribution} that is independent of the source term as the \gls{QNEF} $B_{n}$, which is given by
\begin{equation}
\label{eq:QNEF}
    B_{n} \equiv \left. \dfrac{A_{\text{out}}}{2\omega} \left( \dfrac{dA_{\text{in}}}{d\omega} \right)^{-1}\right|_{\omega = \omega_n}.
\end{equation}
Therefore, the accurate determination of the values of \glspl{QNEF} is crucial in understanding \gls{QNM} contents in the context of \gls{BH} perturbation theory. There are already a large number of literature (e.g., Refs.~\cite{Leaver:1986gd, Glampedakis:2003dn, Berti:2006wq, Dolan:2011fh, Zhang:2013ksa, Oshita:2021iyn, Motohashi:2024fwt, Chen:2024hum}) on \glspl{QNEF} of \glspl{BH} in general relativity and some even beyond \cite{Silva:2024ffz}.

\subsection{This work}
In this paper, we study \gls{QNM} solutions of Kerr \glspl{BH} and their \glspl{QNEF} of perturbation fields with various spin weights ($s=0,\pm 1,\pm 2$) using the \gls{GSN} formalism \cite{Hughes:2000pf, Lo:2023fvv} that extends the original \gls{SN} formalism to work for perturbations with any integer spin weight.
Our work complements previous works mainly in two ways.
First, we use the \gls{GSN} formalism for numerical computations instead of the \gls{MST} method \cite{Zhang:2013ksa, Motohashi:2024fwt} and a Heun-function-based method \cite{Oshita:2021iyn} that are deployed in some recent calculations of \glspl{QNEF} in the literature, allowing us to cross-check existing results. An underappreciated fact is that the \gls{QNEF} defined in Eq.~\eqref{eq:QNEF} under the Teukolsky formalism is actually \emph{not} dimensionless, and the dimension depends on the spin weight $s$. Thus, care must be taken when using results computed with different conventions (e.g., $M=1$ vs $M=1/2$).
Second, we compute both the $+s$ and $-s$ \glspl{QNEF} explicitly for $s = 0,1,2$ \emph{where they are, in fact, different even though they have identical \gls{QNM} spectra}, which is also underappreciated in the literature.
Our novel approach does not involve any tricky manipulation with special functions or further approximations other than integrating the \gls{ODE} numerically and, therefore, can be adopted to solve \gls{QNM} problems (both the frequencies and the corresponding excitation factors) accurately for a wide class of spacetimes.
It performs well even when $|M\omega|$ is large, which is the case when the angular number $\ell$ or the overtone number $n$ is large.

This paper is structured as follows.
In Sec.~\ref{sec:formalism}, we develop the method to numerically solve the source-free \gls{GSN} equation for complex frequencies.
Then, in Sec.~\ref{sec:results}, we specialize to \glspl{QNM}, searching for the allowed complex frequencies and computing their corresponding \glspl{QNEF}.
We discuss our results in Sec.~\ref{sec:Discussions} and summarize our findings in Sec.~\ref{sec:conclusion}.

Throughout this paper, we use geometric units $c=G=M=1$. As a result, the Teukolsky excitation factors reported here coincide with the dimensionless $M^{-2s} B_{n}^{\text{T}, (s)}$.

\section{Generalized Sasaki-Nakamura formalism with complex frequencies \label{sec:formalism}}
Before we embark on the journey of computing \gls{QNM} solutions using the \gls{GSN} formalism, let us first review the essence of the procedure (we refer readers to Ref.~\cite{Lo:2023fvv} for a detailed account of the \gls{GSN} formalism). We then describe how a standard trick in the theory of complex variables (i.e., deforming the integration contour) allows us to numerically solve the \gls{GSN} equation with complex frequencies.

Instead of solving for the Teukolsky function, we solve for the \gls{GSN} function $X(r_*)$, which satisfies the \gls{ODE}
\begin{equation}
\dfrac{d^2 X}{d r_{*}^2} - \mathcal{F} \dfrac{d X}{d r_{*}}	- \mathcal{U} X = 0\,,\label{eq:GSN_equation}
\end{equation}
where the potentials $\mathcal{F}$ and $\mathcal{U}$ are functions of the \gls{BH} parameters, the radial coordinate, the angular numbers characterizing the angular distribution of the field, and the frequency $\omega$. The frequency $\omega$ can be interpreted either physically as the wave frequency near spatial infinity or mathematically as the Fourier variable.

In particular, let us examine the asymptotic behaviors of \gls{GSN} functions as $r_* \in \mathbb{R} \to \pm \infty$. They behave like 
\begin{equation}
X \to \begin{cases}
    \exp\left(\pm i p r_*\right) & \mathrm{as}\; r_* \to -\infty, \\
    \exp\left(\pm i \omega r_*\right) & \mathrm{as}\; r_* \to +\infty, \\	
\end{cases}
\end{equation}
where $p \equiv \omega - ma/(2r_+)$ is the ``effective wave frequency'' at the \gls{BH} horizon.
If the imaginary part of $\omega$ is negative, then $X \sim e^{\pm |\mathrm{Im}\;\omega|r_*}\,e^{\pm i\left(\mathrm{Re}\;\omega\right)r_*}$ as $r_* \to \infty$.
This means that the wave amplitude is either exponentially growing or exponentially decaying, depending on the wave propagation direction. The same phenomenon occurs when $r_* \to -\infty$, with $X \sim e^{\pm |\mathrm{Im}\;\omega|r_*}\,e^{\pm i\left[\left(\mathrm{Re}\;\omega\right) - ma/(2r_+)\right]r_*}$.
This makes accurate numerical integration of the \gls{GSN} equation and the determination of the asymptotic amplitudes particularly challenging, even though the formalism itself works just fine with a complex $\omega$.

In fact, this exponential growth or decay in the amplitude of the \gls{GSN} function is simply a coordinate artifact.
By analytic continuing the \gls{GSN} function $X(r_*)$ from the real line to the entire complex-$r_{*}$ plane and integrating Eq.~\eqref{eq:GSN_equation} along a deformed contour on the complex plane,\footnote{In the context of \gls{BH} perturbation theory, this has already been employed in, for example, Refs.~\cite{1974CMaPh..38...47H, Glampedakis:2003dn}. It has also been adopted to study scattering processes in non-Hermitian quantum mechanics, known as the complex scaling transformation \cite{de1965potential, Moiseyev_2011}, that transforms resonant states into bound states.} one can suppress this coordinate artifact and the \gls{GSN} equation will still admit a plane-wave solution with unit amplitude when $|r_{*}| \to \infty$.\footnote{Another way to suppress this artifact is to use a set of horizon-penetrating hyperboloidally compactified coordinates. See, for example, Ref.~\cite{Ripley:2022ypi}.}

\subsection{Analytic continuation of the Boyer-Lindquist $r$ coordinate}
Recall that the Boyer-Lindquist $r$ coordinate is related to the tortoise coordinate $r_*$ introduced earlier in Eq.~\eqref{eq:rstar_from_r} (for real values of $r$ and $r_{*}$) by
\begin{equation}
\label{eq:drdrstar}
\dfrac{dr}{dr_{*}} \equiv \dfrac{\Delta(r)}{r^2 + a^2}.	
\end{equation}
We shall use this relation and Eq.~\eqref{eq:rstar_from_r} to analytic continue $r(r_{*})$ for complex values of $r_{*}$.
Instead of using the real and the imaginary part of $r_*$, we parametrize $r_{*}$ as
\begin{equation}
\label{eq:contour_rotation}
r_* = r_*^{\rm{mp}} + \rho e^{i\beta},
\end{equation}
where $r_*^{\rm{mp}}$ is a real number and $\rho \in (-\infty, +\infty)$ and $|\beta| < \pi/2$ are two real variables.

\begin{figure}[h]
\includegraphics[scale=0.4]{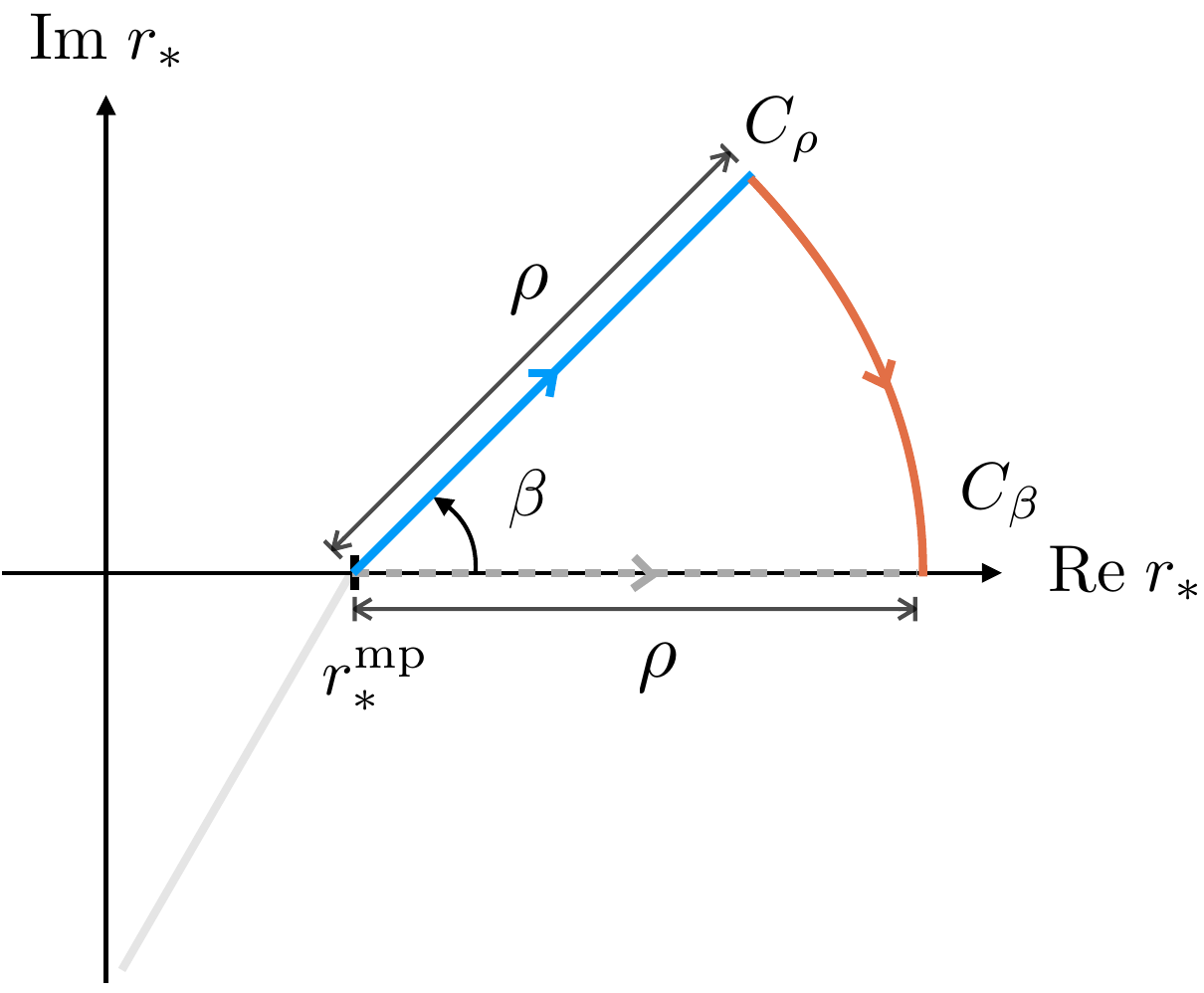}
\caption{\label{fig:integration_contour}Complex-$r_{*}$ plane parametrized by two real variables $\rho$ and $\beta$.
}
\end{figure}

With this parametrization, we can use the notion of a differential in the usual sense, i.e., an infinitesimally small change of a variable. Along a path $C_{\rho}$ with varying $\rho$ but $\beta$ unchanged (say, $\beta = \beta_0$), we have
\begin{equation}
\label{eq:drstardrho}
	dr_* = e^{i\beta_0}d\rho.
\end{equation}
Similarly, along a path $C_{\beta}$ with varying $\beta$ but $\rho$ unchanged (say, $\rho = \rho_0$), we have
\begin{equation}
\label{eq:drstardbeta}
	dr_*  = i\rho_0 e^{i\beta} d\beta.
\end{equation}
For instance, the path $C_{\rho}$ shown in Fig.~\ref{fig:integration_contour} can be parametrized by $r_{*}(\rho)$ with a fixed $\beta$, while the path $C_{\beta}$ in the same figure can be parametrized by $r_{*}(\beta)$ with a fixed $\rho$.

\begin{figure}[h]
\includegraphics[scale=0.6]{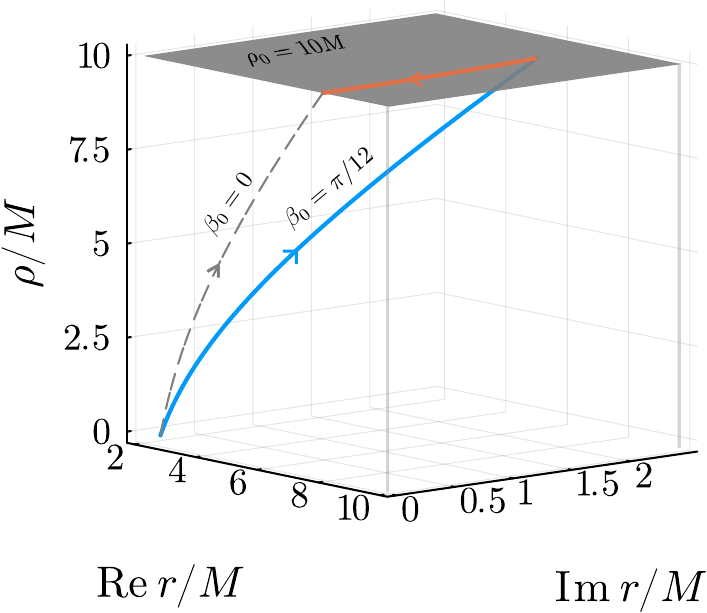}
\caption{\label{fig:rrho}An example of $r(\rho)$ (blue solid curve) and $r(\beta)$ (orange solid curve), respectively, with $r_{*}^{\text{mp}} = 0$, $a = 0$. In particular we are showing $r(\rho)$ with $\beta_0 = \pi/12$ from the matching point $\rho = 0$ up to $\rho = 10$. At $\rho_0 = 10$, $r(\beta)$ connects $r(\rho = 10) \in \mathbb{C}$ with $r(r_* = 10) \in \mathbb{R}$, demonstrating the path independency of $\int dr/dr_{*}$ as required by the analyticity of $r(r_{*} \in \mathbb{C})$.}
\end{figure}
We solve for $r(\rho)$ by integrating the \gls{ODE} obtained from substituting Eq.~\eqref{eq:drstardrho} into Eq.~\eqref{eq:drdrstar}, which is
\begin{equation}
\label{eq:drdrho}
	\dfrac{dr}{d\rho} = e^{i\beta_0} \dfrac{\Delta(r)}{r^2 + a^2},
\end{equation}
together with the initial condition that $r = r(r_{*}^{\text{mp}} \in \mathbb{R})$ given by the inverse of Eq.~\eqref{eq:rstar_from_r}\footnote{Note that $r(r_{*} \in \mathbb{R})$ is multivalued. Here, we choose the branch that maps $r_{*} \in (-\infty, +\infty) \mapsto r \in (r_+, \infty)$.} when $\rho = 0$ (hence the superscript $\text{mp}$, which stands for matching point).
Figure~\ref{fig:rrho} shows an example of $r(\rho)$ (blue solid curve) with $\beta_0 = \pi/12$, $r_{*}^{\text{mp}} = 0$, $a = 0$, and $\rho \in \left[0, 10\right]$.
It diverges from $r(r_{*} \in \mathbb{R})$ (gray dashed curve), i.e., $\beta_0 = 0$, starting at the matching point ($\rho = 0$).
In the same figure, we also show $r(\beta)$ (orange solid curve) with $\rho_0 = 10$ and $\beta \in [0, \pi/12]$. We see that it connects $r(r_{*} = 10e^{i\pi/12})$ to $r(r_{*} = 10)$ as required by the analyticity of $r(r_{*} \in \mathbb{C})$.

\subsection{Analytic continuation of the generalized Sasaki-Nakamura functions\label{subsec:GSN_analytic_continuation}}
Equipped with the analytic-continued $r(r_{*})$, we can analytic continue also the \gls{GSN} potentials $\mathcal{F}(r_*)$ and $\mathcal{U}(r_*)$ by evaluating these potentials at complex values of $r = r(r_{*} \in \mathbb{C})$.\footnote{Note that these functions have poles located at $r = \pm ia$. Since these poles are not in the image of $r(r_*)$, those functions are still analytic in $r_*$.}
Following that, we can analytic continue GSN functions $X(r_*)$ by integrating Eq.~\eqref{eq:GSN_equation}.
Since numerically it is easier to integrate an \gls{ODE} of a real variable instead of a complex one, we adopt the parametrization in Eq.~\eqref{eq:contour_rotation} and vary $\rho$ or $\beta$ at a time.

\subsubsection{Varying $\rho$}
Along the $C_{\rho}$ path, $\rho$ is the variable with $\beta = \beta_0$ held fixed.
Substituting Eq.~\eqref{eq:drstardrho} into Eq.~\eqref{eq:GSN_equation}, we have
\begin{equation}
\label{eq:GSN_along_rho}
	\dfrac{d^2 X}{d \rho^2} - e^{i\beta_0}\mathcal{F} \dfrac{d X}{d \rho} - e^{2i\beta_0} \mathcal{U} X = 0,
\end{equation}
which is different from the original \gls{GSN} equation. However, $X(\rho)$ is still a solution to the original \gls{GSN} equation evaluated at $X(r_* = r_*^{\rm{mp}} + \rho e^{i\beta_0})$.

Notice that Eq.~\eqref{eq:GSN_along_rho} is almost identical to the original \gls{GSN} equation in Eq.~\eqref{eq:GSN_equation} by rescaling the potentials as
\begin{align}
	\tilde{\mathcal{F}}(\rho) & \equiv e^{i\beta}\mathcal{F}(r(\rho)),\\
	\tilde{\mathcal{U}}(\rho) & \equiv e^{2i\beta} \mathcal{U}(r(\rho)).
\end{align}
Therefore, the transformed \gls{GSN} equation can be written as
\begin{equation}
	\dfrac{d^2 X}{d \rho^2} - \tilde{\mathcal{F}} \dfrac{d X}{d \rho} - \tilde{\mathcal{U}} X = 0.
\end{equation}
We can also reuse \emph{exactly} the \textit{Ans\"atze} constructed for the original \gls{GSN} equation in Ref.~\cite{Lo:2023fvv} to help solve this new equation. Note that we will need $\left\{ X, dX/d\rho \right\}$ instead of $\left\{ X, dX/dr_* \right\}$, but the derivative $dX/d\rho$ is related to $dX/dr_*$ by
\begin{equation}
\begin{aligned}
	\left.\dfrac{dX}{d\rho}\right|_{\rho=\rho_0}  & = \dfrac{dr_*}{d\rho} \left.\dfrac{dX}{dr_*}\right|_{r_* = r_*^{\rm{mp}} + \rho_0 e^{i\beta}} \\
	& = e^{i\beta} \left.\dfrac{dX}{dr_*}\right|_{r_* = r_*^{\rm{mp}} + \rho_0 e^{i\beta}},
\end{aligned}
\end{equation}
which is nothing but the chain rule written out explicitly.

Asymptotically as $\rho \to \pm \infty$, we see that the solutions $\tilde{X}$ behave as
\begin{equation}
X(\rho) \to \begin{cases}
    \exp\left[\pm i \left(e^{i\beta} p\right) \rho\right] & \mathrm{as}\; \rho \to -\infty \\
    \exp\left[\pm i \left(e^{i\beta} \omega\right) \rho\right] & \mathrm{as}\; \rho \to +\infty
\end{cases}.
\end{equation}
If we choose $\beta$ carefully such that the combination $e^{i\beta} \omega$ or $e^{i\beta}p$ is a purely real number, then asymptotically $X(\rho)$ will behave like a plane wave of constant amplitude even with a complex $\omega$.

Recall that we can express any complex number $k$ as
\begin{equation}
	k = |k|e^{i\arg k}.	
\end{equation}
Therefore, if we choose $\beta = -\arg k$, where $k$ can be $\omega$ or $p$ as appropriate, then
\begin{equation}
	e^{i\beta}k = e^{-i\arg k}k = (k^{*}/|k|)k = |k|^2/|k| = |k|,
\end{equation}
where a superscripted asterisk denotes the complex conjugate, is indeed a purely real number.

Since $p$ coincides with $\omega$ only when $m = 0$ or $a = 0$, we need to deform the integration contour into two broken line segments (i.e., with two different slopes, given precisely by $\beta$) in the complex-$r_{*}$ plane in generic cases.
The appropriate choices of $\beta$ are as follows---when $\rho < 0$,
\begin{equation}
\label{eq:beta_negrho}
\beta = \begin{cases}
	-\arg p & \, \text{when } \text{Re }p > 0, \\
	-\arg p + \pi & \, \text{when } \text{Re }p < 0, \\
\end{cases}
\end{equation}
and when $\rho > 0$, we have
\begin{equation}
\label{eq:beta_posrho}
\beta = \begin{cases}
	-\arg \omega & \, \text{when } \text{Re }\omega > 0, \\
	-\arg \omega + \pi & \, \text{when } \text{Re }\omega < 0. \\
\end{cases}
\end{equation}
In particular, $\beta$ is always within the allowed range $\left( -\pi/2, \pi/2 \right)$ with these choices.\footnote{These choices, therefore, do not work for purely imaginary modes (e.g., see Ref.~\cite{Cook:2016ngj}) where the real part of $\omega$ or $p$ vanishes as $\beta$ would be $\pm \pi/2$.}
Note that adding an extra $\pi$ to $\beta$ is identical to flipping $\rho \mapsto -\rho$ as $e^{i\pi} = -1$.

\subsubsection{Varying $\beta$}
Along the $C_{\beta}$ path, $\beta$ is the variable with $\rho = \rho_0$ held fixed instead.
Substituting Eq.~\eqref{eq:drstardbeta} into Eq.~\eqref{eq:GSN_equation}, the new \gls{ODE} that needs to be solved reads
\begin{equation}
\label{eq:GSN_along_beta}
	\dfrac{d^2 X}{d\beta^2} - i \left( 1 + \rho_0 e^{i\beta} \mathcal{F}\right) \dfrac{dX}{d\beta} + \left( \rho_0 e^{i\beta} \right)^2\mathcal{U} X = 0.
\end{equation}
The derivative $dX/d\beta$ is \emph{no longer} a simple constant rescaling of $dX/dr_*$ compared to $dX/d\rho$.
In particular, when $|\rho_0| \to \infty$, the \gls{ODE} simplifies drastically as the potentials approach
\begin{equation}
\label{eq:asymptotic_potentials}
\begin{aligned}
\mathcal{F} & \to \;\;\;\;\; 0 \;\;\;\;\;\;\, \text{as } \rho \to \pm \infty,\\ 
\mathcal{U} & \to \begin{cases}
-p^2 & \text{as } \rho \to -\infty, \\
-\omega^2 & \text{as } \rho \to +\infty.
\end{cases}
\end{aligned}
\end{equation}
The \gls{ODE} can be solved analytically in this limit as
\begin{equation}
\label{eq:X_beta_exact}
	X(\beta) = c_1 \exp\left(ik\rho_0 e^{i\beta}\right) + c_2 \exp\left(-ik\rho_0 e^{i\beta}\right),
\end{equation}
where $c_{1,2}$ are some constants to be determined by the initial conditions at $\beta = \beta_0$ and that $k = \omega$ when $\rho_0 \to \infty$ and $k = p$ when $\rho_0 \to -\infty$.
For example, when $\rho_0 \to \infty$, we have
\begin{equation}
    X(\rho_0 \to \infty, \beta_0) = c_1 \exp\left(i\omega \rho_0 e^{i\beta_0}\right) + c_2 \exp\left(-i\omega\rho_0 e^{i\beta_0}\right).
\end{equation}

When $\beta = 0$, i.e., $r_* = r_*^{\rm mp} + \rho_0$ is purely real, the solution becomes
\begin{equation}
\label{eq:Ain_Aout_from_c1_c2}
\begin{aligned}
	X(\rho \to \pm \infty, \beta = 0) & = c_1 \exp \left[ ik\left(r_* - r_*^{\rm mp}\right)  \right] \\
	& + c_2 \exp \left[ -ik\left(r_* - r_*^{\rm mp}\right)  \right] \\
	& = \underbrace{c_1 e^{-ik r_*^{\rm mp}}}_{A_{\rm out}} e^{ikr_*} + \underbrace{c_2 e^{ikr_*^{\rm mp}}}_{A_{\rm in}} e^{-ikr_*}.
\end{aligned}
\end{equation}
This is a key result---it means that we can extract the asymptotic amplitudes $A_{\text{in}}$ and $A_{\text{out}}$ of $X(r_*)$ directly from the asymptotic amplitudes $c_{1}$ and $c_{2}$ of $X(\rho)$, where $X(\rho)$ can be solved accurately (see, for example, Ref.~\cite{Lo:2023fvv}) with an off-the-shelf numerical integrator.

We now have everything needed to compute $A_{\text{in}}(\omega)$ and $A_{\text{out}}(\omega)$ for a complex $\omega$, where a Kerr \gls{QNM} is a solution to the \gls{GSN} equation in Eq.~\eqref{eq:GSN_equation} with a vanishing $A_{\text{in}}$.

\section{\label{sec:results}Results}
We are now ready to compute \gls{QNM} solutions of Kerr \glspl{BH} using the \gls{GSN} formalism we have just developed.
In Sec.~\ref{subsec:complex_freq_soln}, we first solve $X_{\text{in}}$ and $X_{\text{up}}$ [cf. Eqs.~\eqref{eq:u_in} and~\eqref{eq:u_up}] for some complex frequencies.
Then, in Sec.~\ref{subsec:QNM_freq}, we search for \gls{QNM} frequencies $\omega_{\text{QNM}}$ recovering previous results~\cite{Leaver:1985ax, Onozawa:1996ux, Cook:2014cta,Berti:2009kk} while extending most of them to higher \gls{BH} rotation.
Using these values of the \gls{QNM} frequencies, in Sec.~\ref{subsec:QNEF}, we compute the \gls{QNEF} for each of those modes as defined in Eq.~\eqref{eq:QNEF}.

Numerically, the radial Teukolsky and the \gls{GSN} equations are solved with a \texttt{julia} \cite{Julia-2017} implementation of the \gls{GSN} formalism in \texttt{GeneralizedSasakiNakamura.jl} \cite{Lo:2023fvv}.\footnote{\url{https://github.com/ricokaloklo/GeneralizedSasakiNakamura.jl} from \texttt{v0.6.0} onward.}
In particular, they are solved with the Vern9 algorithm \cite{2010NuAlg..53..383V} implemented in \texttt{DifferentialEquations.jl} \cite{rackauckas2017differentialequations} and double-precision floating-point numbers. %
As for the angular Teukolsky equation, we use a spectral decomposition method \cite{Hughes:1999bq, Cook:2014cta} implemented in \texttt{SpinWeightedSpheroidalHarmonics.jl} \cite{Lo:2023fvv},\footnote{\url{https://github.com/ricokaloklo/SpinWeightedSpheroidalHarmonics.jl}.}
which works with complex frequencies.
In particular, we optimize $r_*^{\rm mp}$ [cf.~Eq.~\eqref{eq:contour_rotation}] for each mode such that the matching point is as close to the horizon as possible for a better computation efficiency (see the \hyperref[app:choosing_mp]{Appendix}).

\subsection{Solutions to the generalized Sasaki-Nakamura equation  \label{subsec:complex_freq_soln}}

To obtain $X_{\text{in}}$, for example, we integrate Eq.~\eqref{eq:GSN_along_rho} from $r_{*}^{\text{in}} = r_{*}^{\text{mp}} + \rho_{\text{in}}e^{i\beta(\rho_{\text{in}})}$, where $\rho_{\text{in}} < 0$ and $\beta(\rho_{\text{in}})$ is determined by Eq.~\eqref{eq:beta_negrho}, to $\rho = 0$ (which is at $r_{*} = r_{*}^{\text{mp}}$) with $\beta$ held fixed. 
We then continue the integration from $\rho = 0$ to $\rho = \rho_{\text{out}} > 0$ [correspondingly, $r_{*} = r_{*}^{\text{mp}} + \rho_{\text{out}}e^{i\beta(\rho_{\text{out}})}$] along a \emph{different} constant-$\beta$ direction, where $\beta = \beta(\rho_{\text{out}})$ is determined by Eq.~\eqref{eq:beta_posrho}.
Similarly for $X_{\text{up}}$, we integrate Eq.~\eqref{eq:GSN_along_rho} from $\rho_{\text{out}} > 0$ to 0 with $\beta = \beta(\rho_{\text{out}})$ and then from 0 to $\rho_{\text{in}} < 0$ with $\beta = \beta(\rho_{\text{in}})$.

\begin{figure}[h]
\centering
\includegraphics[width=\columnwidth]{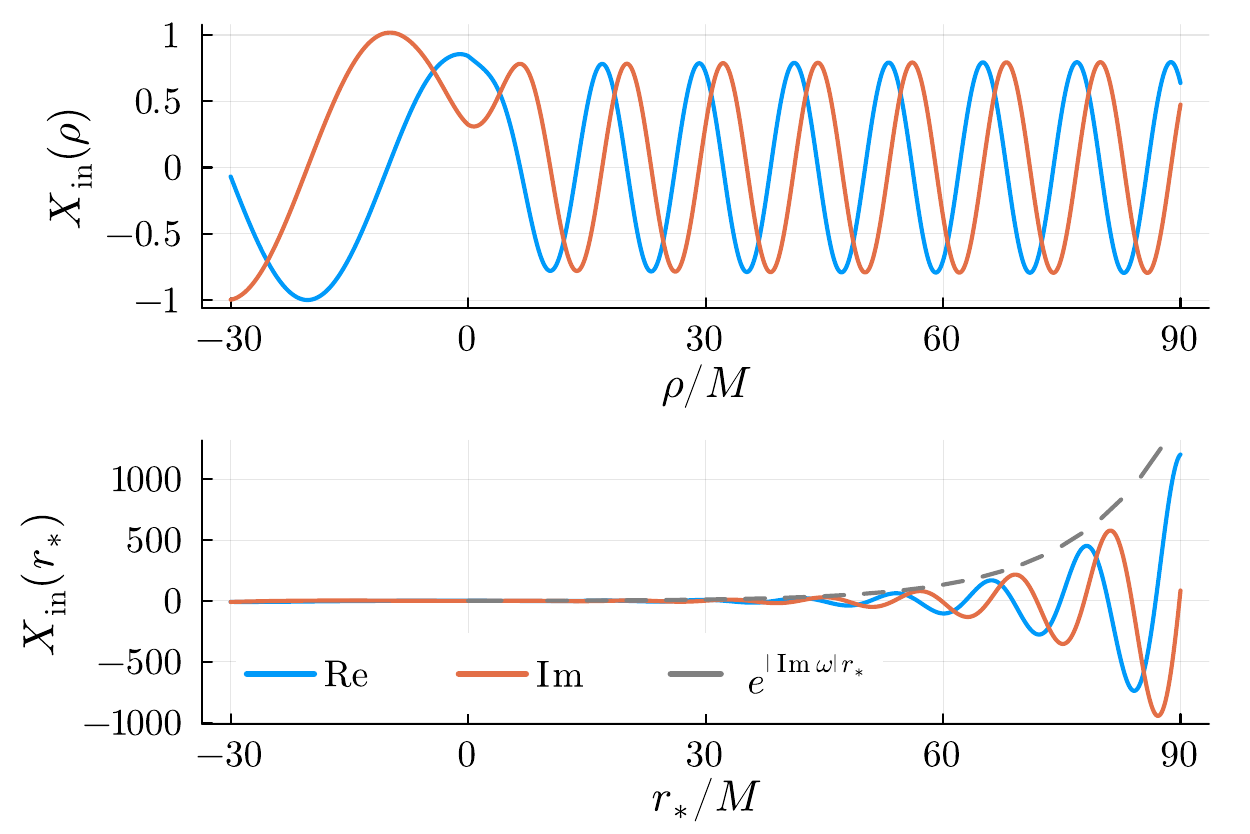}
\caption{\label{fig:Xin}\gls{GSN} solution $X_{\text{in}}$ with $s=-2, \ell=m=2$, $a/M=0.68$, and $M\omega = 0.5239751-0.0815126i$ along a deformed integration contour parametrized by $\rho$ (upper panel) and along the real axis of $r_{*}$ (lower panel), respectively.}
\end{figure}

An example of $X_{\text{in}}$ with $s=-2, \ell=m=2$, $a=0.68$, and $\omega = 0.5239751-0.0815126i$ is shown in Fig.~\ref{fig:Xin}.
As expected, $X_{\text{in}}(\rho)$ in the upper panel behaves like a plane wave near the two boundaries, while $X_{\text{in}}(r_*)$ in the lower panel exhibits an exponential blowup in the amplitude near $r_* \to \infty$, following the expected $e^{|\mathrm{Im}\;\omega| r_*}$ scaling (gray dashed line).

In theory, for each value of $\rho = \rho_0$, we integrate Eq.~\eqref{eq:GSN_along_beta} from $\beta = \beta(\rho_0)$ back to $\beta = 0$ on the real axis.
In practice, we are mostly interested in the asymptotic values when $r_{*} \to \pm \infty$. In this case, Eq.~\eqref{eq:X_beta_exact} gives the desired solution with two constants $c_{1,2}$ determined by the asymptotic solutions $X_{\text{in}}(\rho)$.
We follow the procedures given in Sec.~III~B in Ref.~\cite{Lo:2023fvv}, i.e., matching the numerical solution of $X_{\text{in}}$ that was solved up to a numerical outer boundary at $\rho = \rho_{\text{out}}$ with an analytical \textit{Ansatz}, to extract the values of $c_{1,2}$ and then $A_{\text{out}}$ and $A_{\text{in}}$ using Eq.~\eqref{eq:Ain_Aout_from_c1_c2}.

For instance, with the example shown in Fig.~\ref{fig:Xin}, the reflection amplitude $A_{\text{out}}$ is about $-0.801 + 0.0291i$,\footnote{We are showing these numbers to three significant figures only for demonstration purposes.} while the incidence amplitude $A_{\text{in}}$ is about $-1.51 \times 10^{-7} - 7.98 \times 10^{-8}i$.
It is no coincidence that $A_{\text{in}}$ here is tiny---this particular frequency was chosen to be very close to the fundamental mode for the $|s|=\ell=m=2$ \gls{QNM}.

\subsection{Quasinormal mode frequencies \label{subsec:QNM_freq}}
To establish the \gls{QNM} frequency spectrum for a perturbation field of spin weight $s$ and angular mode labeled by $\left(\ell,m\right)$, we solve the nonlinear equation $A_{\text{in}}(\omega_{\text{QNM}}) = 0$ for $a/M \in \left[0, 1\right)$ using the Newton-Raphson method \cite{10.5555/1403886}.
While verifying a \gls{QNM} frequency can be done very quickly,
the tabulation of those frequencies for different $s \ell m$ modes, on the other hand, is a time-consuming task.

Fortunately, one can save computational time by exploiting symmetries and identities of homogeneous solutions to the Teukolsky equation. For instance, perturbations of spin weight $s$ and $-s$ share the same \gls{QNM} frequency spectrum.
This is because asymptotic amplitudes of a homogeneous solution with spin weight $s$ are proportional to that of $s \mapsto -s$ by the virtue of the Teukolsky-Starobinsky identities \cite{Starobinskil:1974nkd, Teukolsky:1974yv}.
Therefore, when tabulating \gls{QNM} frequencies, one needs only to compute values with either $s$ or $-s$.

Another useful relation is that
\begin{equation}
\label{eq:mirror_symmetry}
	{}_{s}u_{\ell,m,-\omega^{*}} = {}_{s}u_{\ell,-m,\omega}^{*},
\end{equation}
where $u$ can be either a homogeneous solution to the radial Teukolsky equation, $R$, or that to the \gls{GSN} equation, $X$.
This implies that for a given $s \ell m$ mode, one needs only to compute \gls{QNM} frequencies, for example, with a positive real part, and the rest of the spectrum can be ``mirrored'' (flipping only the sign of the real part) from the corresponding $-m$ mode. 
More explicitly, if we denote those \gls{QNM} frequencies with a positive real part as $\{ \omega^{+}_{s \ell m n} \}$, which are sometimes referred to as ``ordinary'' modes, then there is also another set of \gls{QNM} frequencies given by $\left\{ \omega^{-}_{s \ell m n} \equiv -{\omega_{s \ell (-m) n}^{+}}^{*} \right\}$, which are also referred to as mirror modes in the literature.

In this work, we tabulate \gls{QNM} frequencies from the nonspinning limit with $a = 0$ to the near-extremal limit with $a = 1 - 10^{-4}$, and $|s| \leq \ell \leq 7$, $-\ell \leq m \leq \ell$, $n \leq 3$ for both scalar\footnote{Except for $s = \ell = m = 0$, where we tabulate only the $n = 0, 1$ modes.} and electromagnetic \glspl{QNM} (i.e., $s = 0, \pm 1$, respectively) and $n \leq 7$ for gravitational \glspl{QNM} (i.e., $s = \pm 2$). We explicitly compute only \gls{QNM} frequencies where their real part are positive.
These tables are available online \cite{qnmqnefdata}.

\begin{figure}[h]
\includegraphics[width=\columnwidth]{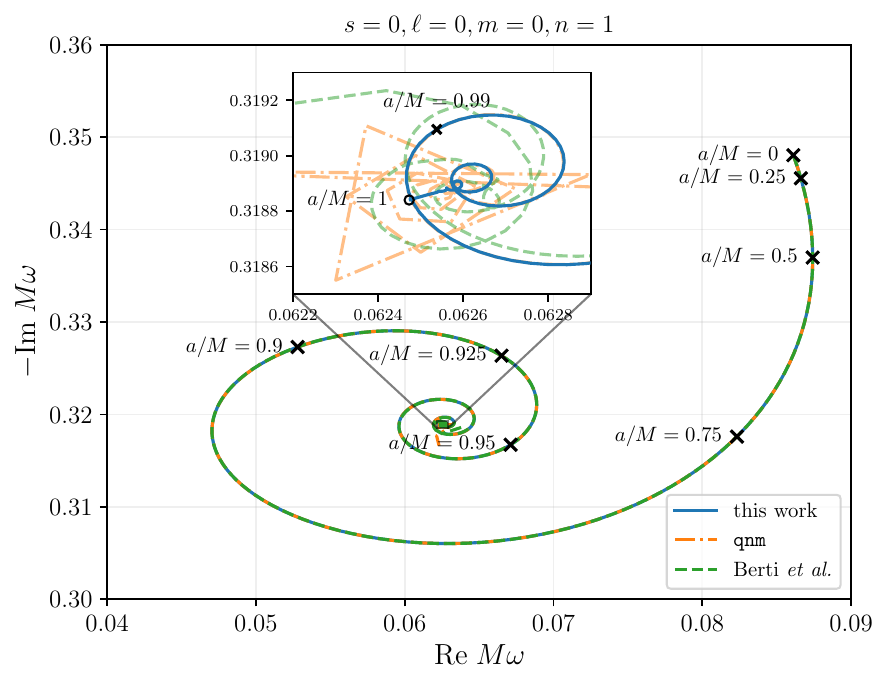}
\caption{\label{fig:s0l0m0n1}Trajectory of the $s = 0, \ell = 0, m = 0$, and $n = 1$ \gls{QNM} frequencies. Discrepancies of the frequencies from this work with literature begin only at near-extremal spins where $a/M \gtrsim 0.99$. Our trajectory of the \gls{QNM} (blue solid curve), computed up to $a/M = 0.9999$, approaches to the extremal $a = M$ limit (open circle) found in the literature \cite{Richartz:2015saa}.}
\end{figure}

It is important to note that most of the \glspl{QNM} tabulated in this work have been previously computed and that our main goal here is to cross-check the frequency calculations.
As an example, Fig.~\ref{fig:s0l0m0n1} shows the trajectory in the complex plane, parametrized by $a$, of the $s = \ell = m = 0, n = 1$ \gls{QNM} frequencies obtained in this work (blue), using the \texttt{qnm} package \cite{Stein:2019mop} (orange) and from Berti \etal{} \cite{Berti:2005ys, Berti:2009kk} (green), respectively.\footnote{Note that the calculations of Kerr \gls{QNM} frequencies in Refs.~\cite{Berti:2005ys, Berti:2009kk}, and available online~\cite{BertiCardoso}, are unreliable very close to the Kerr extremal limit (roughly, when $a \gtrsim 0.999$).}
The three trajectories agree well for most values of $a$. Discrepancies begin only at near-extremal spins $a \gtrsim 0.99$, where our calculation suggests that the \gls{QNM} frequency approaches steadily to $\omega \approx 0.0625 - 0.3188i$ at $a = 0.9999$, which matches with the value presented in Table I in Ref.~\cite{Richartz:2015saa} for the extremal $a = 1$ case (open circle in the inset in Fig.~\ref{fig:s0l0m0n1}) when rounded to the same precision. On the other hand, the trajectories from the literature show chaotic behaviors, suggesting that these are numerical artifacts in their calculations.

\begin{figure*}
\centering
\vspace{5em}
\includegraphics[width=2\columnwidth]{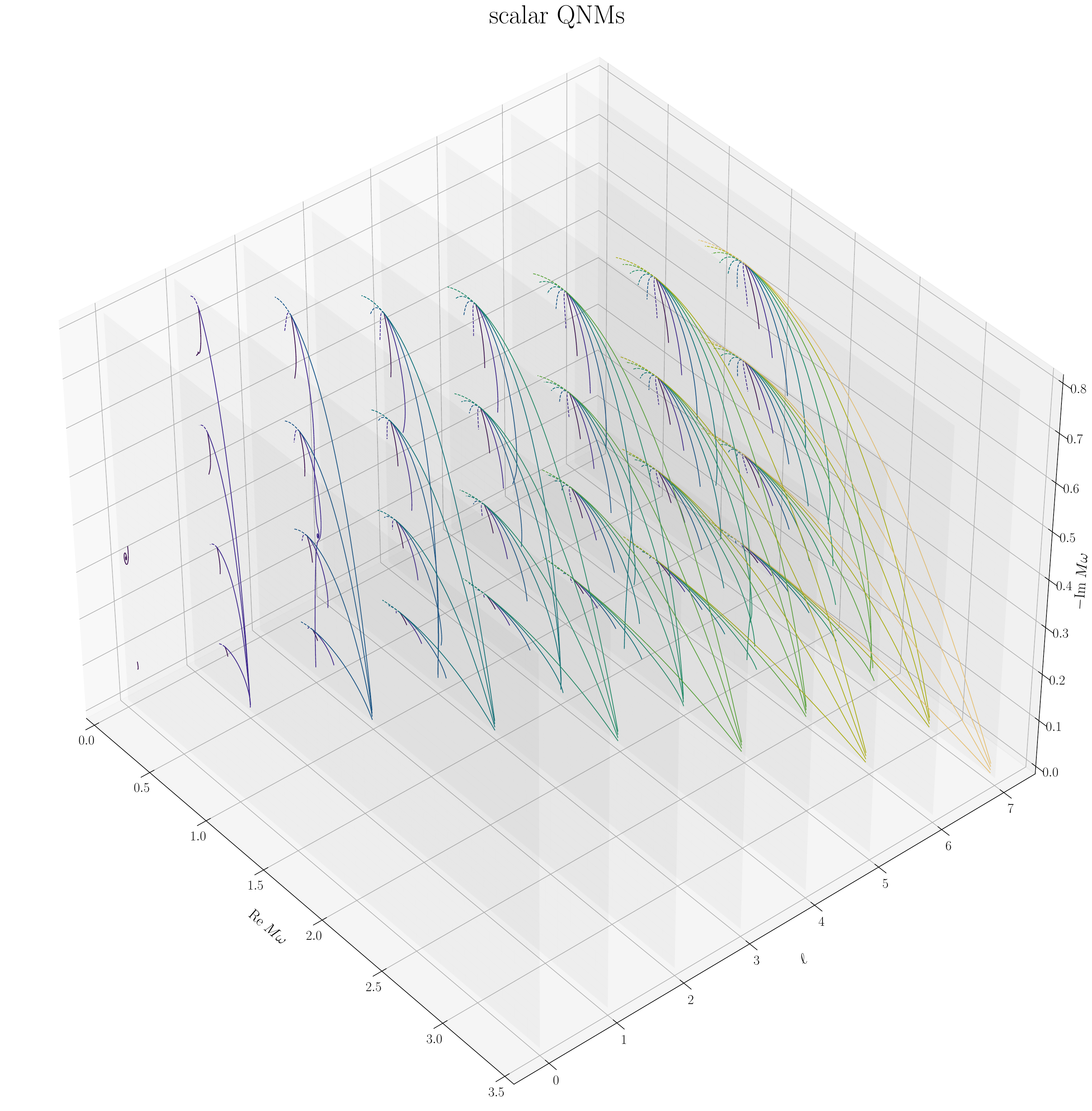}
\caption{\label{fig:scalar_qnm}
Trajectories of scalar ($s = 0$) \glspl{QNM} tabulated in this work ($-\ell \leq m \leq \ell$ for $0 \leq \ell \leq 7$) parametrized by $a/M \in [0, 1 - 10^{-4}]$, up to their third overtones (i.e., $n \leq 3$), except for the $s = \ell = m = 0$ mode. \plotinstr{}}
\end{figure*}
\begin{figure*}
\centering
\vspace{7em}
\includegraphics[width=2\columnwidth]{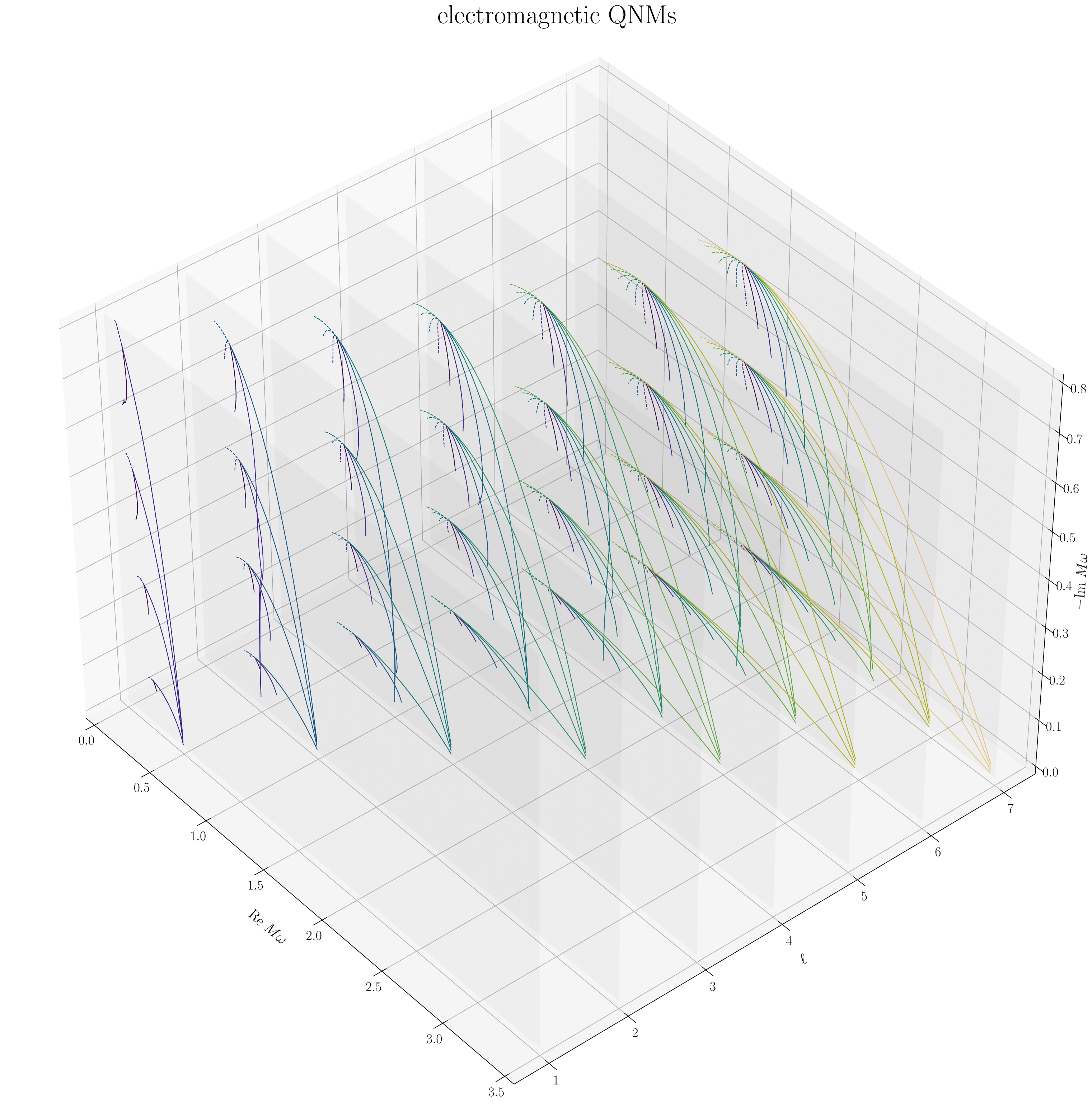}
\caption{\label{fig:electromagnetic_qnm}The same as Fig.~\ref{fig:scalar_qnm}, but for electromagnetic ($s = \pm 1$) \glspl{QNM} tabulated in this work ($-\ell \leq m \leq \ell$ for $1 \leq \ell \leq 7$), up to their third overtones (i.e., $n \leq 3$). Numerical values of these \gls{QNM} frequencies are available online \cite{qnmqnefdata}.}
\vspace{7em}
\end{figure*}
\begin{figure*}
\centering
\vspace{7em}
\includegraphics[width=2\columnwidth]{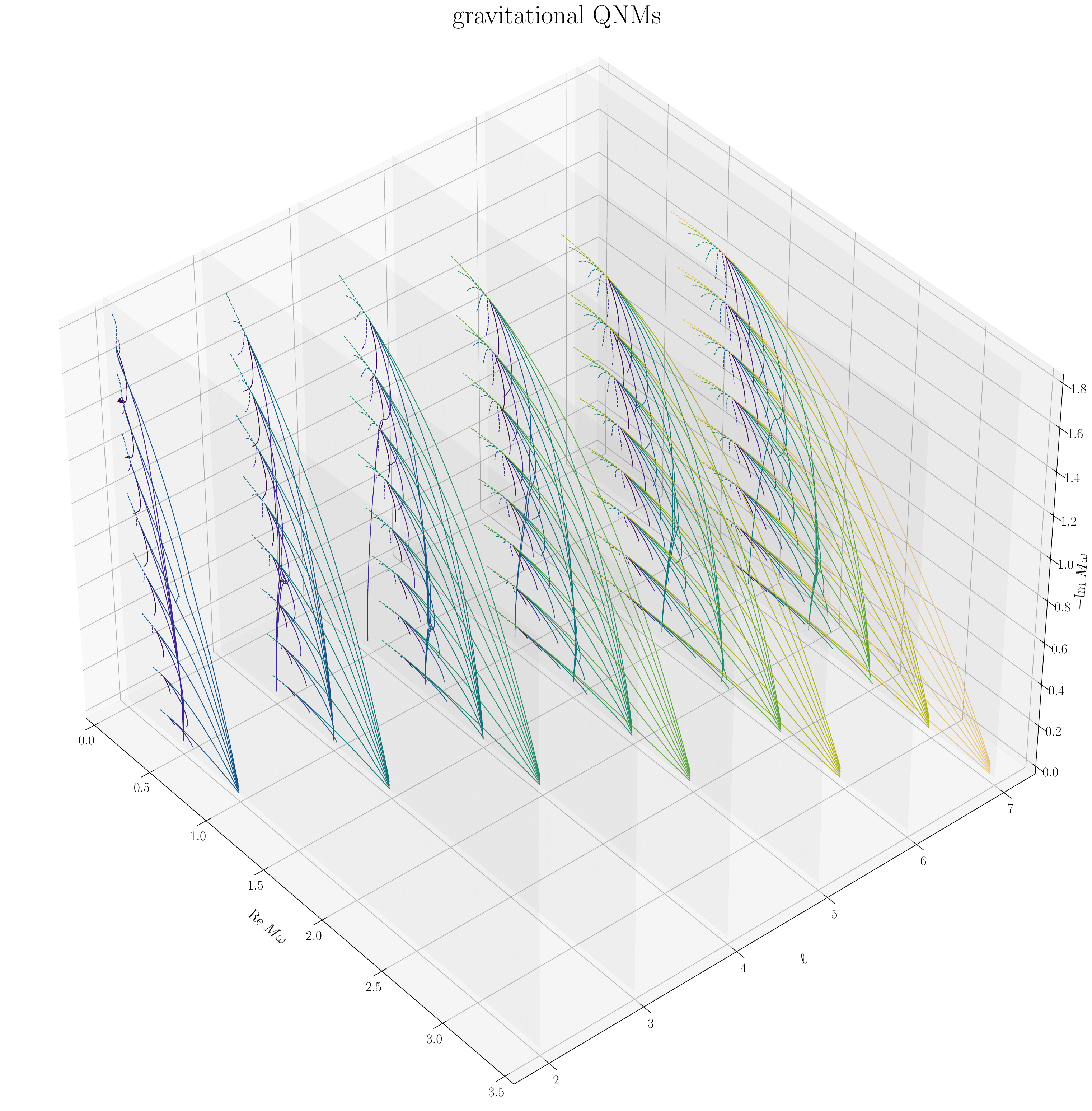}
\caption{\label{fig:gravitational_qnm}The same as Fig.~\ref{fig:scalar_qnm}, but for gravitational ($s = \pm 2$) \glspl{QNM} tabulated in this work ($-\ell \leq m \leq \ell$ for $2 \leq \ell \leq 7$), up to their seventh overtones (i.e., $n \leq 7$). Numerical values of these \gls{QNM} frequencies are available online \cite{qnmqnefdata}.}
\vspace{7em}
\end{figure*}

Figures~\ref{fig:scalar_qnm}--\ref{fig:gravitational_qnm} show the trajectories of all the scalar, electromagnetic, and gravitational \glspl{QNM} tabulated in this work, respectively.
In all these figures, $+|m|$ and $m = 0$ modes are plotted with solid lines, while $-|m|$ modes are plotted with dashed lines. They are connected at $a = 0$ where the \gls{QNM} frequencies do not depend on $m$.
Trajectories with the same value of $|m|$ are plotted with the same color.

Note that these plots are intended to help understand general trends and anomalies in those \gls{QNM} trajectories and inspect our calculations for numerical artifacts. In Sec.~\ref{sec:Discussions}, we will discuss in depth some of their features.

\subsection{Quasinormal mode excitation factors \label{subsec:QNEF}}
Once the frequency of a \gls{QNM} is determined, it is straightforward to compute its corresponding \gls{QNEF} using Eq.~\eqref{eq:QNEF}, as $A_{\text{in}}$ and $A_{\text{out}}$ have already been computed during the process.
In particular, since $A_{\text{in}}(\omega)$ is analytic in $\omega$ \cite{1974CMaPh..38...47H}, therefore $dA_{\text{in}}/d\omega$ can be computed either along the real or the imaginary axis of $\omega$, and both approaches should give the same result.
Here, we use a central finite differencing scheme with Richardson extrapolation \cite{RIDDERS198275} to estimate the derivative numerically.

\glspl{QNEF} can be defined with either the Teukolsky amplitudes or the \gls{SN} amplitudes by using the appropriate $A_{\text{in, out}}^{\text{T/SN}}$ in Eq.~\eqref{eq:QNEF}. A Teukolsky \gls{QNEF} $B_{n}^{\text{T}}$ and the corresponding \gls{SN} \gls{QNEF} $B_{n}^{\text{SN}}$ are off by some known frequency-dependent conversion factors where
\begin{equation}
    B_{n}^{\text{T}} = \left( \dfrac{B^{\text{ref}}_{\text{T}}}{B^{\text{ref}}_{\text{SN}}} \right) \left( \dfrac{B^{\text{inc}}_{\text{T}}}{B^{\text{inc}}_{\text{SN}}} \right)^{-1} B_{n}^{\text{SN}},
\end{equation}
and the expressions for $B^{\text{ref}}_{\text{T}}/B^{\text{ref}}_{\text{SN}}$ and $B^{\text{inc}}_{\text{T}}/B^{\text{inc}}_{\text{SN}}$ can be found in, for example, Ref.~\cite{Lo:2023fvv}.

A much-underappreciated fact regarding \glspl{QNEF} is that perturbations of spin $s$ and $-s$ (with $s \neq 0$) have \emph{different} \glspl{QNEF} despite sharing the same frequency spectrum. This can be seen from the Teukolsky-Starobinsky identities \cite{Starobinskil:1974nkd, Teukolsky:1974yv} mentioned earlier in the text.
For $s = \pm 2$, we have 
\begin{equation}
\label{eq:Bpm2conv}
	B^{\text{T},(+2)}_{n} = \dfrac{|\mathcal{C}|^2}{256\omega^8} B^{\text{T},(-2)}_{n},
\end{equation}
where $B^{\text{T},(\pm 2)}_{n} $ refers to the Teukolsky \gls{QNEF} for $s=\pm 2$, respectively, and that the expression for $|\mathcal{C}|^2$ can be found in Ref.~\cite{Teukolsky:1974yv}.
Despite what the notation might suggest, $|\mathcal{C}|^2$ can actually be complex  \cite{Casals:2020fsb}.
Similarly, for $s = \pm 1$, we have
\begin{equation}
\label{eq:Bpm1conv}
    B^{\text{T},(+1)}_{n} = \dfrac{\mathcal{B}^2}{16\omega^4} B^{\text{T},(-1)}_{n},
\end{equation}
where the expression for $\mathcal{B}^2$ can again be found in Ref.~\cite{Teukolsky:1974yv}.

Here, we tabulate the \glspl{QNEF} for the \glspl{QNM} shown in Sec.~\ref{subsec:QNM_freq}.
In this work, we compute $B^{(\pm s)}_{n}$ explicitly and do not use Eqs.~\eqref{eq:Bpm2conv} and \eqref{eq:Bpm1conv} to obtain $B^{(+s)}_{n}$ from $B^{(-s)}_{n}$ and vice versa.
This is because the conversion factors involve large powers of $\omega$ (more concretely, $\omega^{8}$ for $s = \pm 2$ and $\omega^{4}$ for $s = \pm 1$), and, thus, the numerical values of $\omega$ need to be determined to higher precision in order to convert $B^{(\pm s)}_{n}$ from $B^{(\mp s)}_{n}$ computed numerically. %
Moreover, we tabulate only the \glspl{QNEF} for ``ordinary modes'' ($\omega_{s\ell mn}^{+}$, where the real part of the frequencies are positive), and the \glspl{QNEF} for mirror modes $\omega_{s\ell mn}^{-}$ are related to their ordinary mode counterparts simply by a complex conjugation because of Eq.~\eqref{eq:mirror_symmetry}, i.e.,
\begin{equation}
	\left. B_{n} \right|_{\omega = \omega_{s\ell m n}^{-}} =  \left. \left. B_{n} \right|_{\omega = \omega_{s \ell (-m) n}^{+} } \right.^{*}.
\end{equation}
These tables are available online \cite{qnmqnefdata}, and a gallery of plots showing the trajectories of the mode frequency and the corresponding \gls{QNEF} for all the \glspl{QNM} tabulated in this work can be found online \cite{qnminteractiveplot}.

It is again important to note that most of the \glspl{QNEF} tabulated in this work have been previously computed.
As an example, Table~\ref{tab:gravitational_QNEF} shows a few gravitational ($s = \pm 2$) \glspl{QNEF} from this work and Motohashi \cite{Motohashi:2024fwt, motohashi_2025_14380191}, respectively. The two works agree while using completely different and independent approaches.
Note that Refs.~\cite{Motohashi:2024fwt, motohashi_2025_14380191} use a convention $M = 1/2$ instead of $M = 1$.
The scaling with $M$ for Teukolsky \glspl{QNEF} is nontrivial and $s$ dependent.
In fact, the combination $M^{-2s} B_{n}^{\text{T}, (s)}$ gives the dimensionless Teukolsky \glspl{QNEF}, while \gls{SN} \glspl{QNEF} are trivially dimensionless.\footnote{This is because the amplitudes for the purely ingoing and the purely outgoing part of the Teukolsky solution $R_{\text{in}}(r \to \infty)$ near infinity, i.e., $u_{\text{in}}^{\infty}$ and $u_{\text{out}}^{\infty}$ in Eq.~\eqref{eq:u_in}, respectively, scale differently with $r$ depending on the value of $s$, while those for the \gls{SN} solution $X_{\text{in}}(r \to \infty)$ do not scale with $r$ at all.}
For example, one would need to divide the values for the gravitational \glspl{QNEF} from Refs.~\cite{Motohashi:2024fwt, motohashi_2025_14380191} by $16$ to convert to the $M = 1$ convention properly. 

\begin{table*}[ht!]
\caption{\label{tab:gravitational_QNEF}A table of a few gravitational \glspl{QNEF}, rounded to the nearest sixth decimal place. The full tables for scalar, electromagnetic, and gravitational \glspl{QNEF} are available online \cite{qnmqnefdata}.}
\begin{ruledtabular}
\begin{tabular}{cccdccc}
 & & & & $M = 1$ & $M = 1/2$ & $M = 1$\\
\cmidrule(lr){5-7}
$\ell$ & $m$ & $n$ & \multicolumn{1}{c}{$a/M$} & $B_{n}^{\text{T}, (-2)}$ (this work) & $1/16 \times ( B_{n}^{\text{T}, (-2)}$ from Refs.~\cite{Motohashi:2024fwt, motohashi_2025_14380191}$)$ & $B_{n}^{\text{T}, (+2)}$ (this work) \\
 &  &  & 0 & $\;\;\;0.001599 - 0.001055i$ & $\;\;\;0.001599 - 0.001055i$ & $\;\;\;2.778095 + 8.976020i$ \\
2 & 2 & 0 & 0.7 & $\;\;\;0.002967 + 0.010638i$ & $\;\;\;0.002967 + 0.010638i$ & $-1.268880 + 0.786292i$ \\
 &  &  & 0.9999 & $\;\;\;0.000124 + 0.005321i$ & $\;\;\;0.000124 + 0.005321i$ & $-0.000005 + 0.004066i$ \\
\\
 & & & 0 & $\;\;\;0.006990 + 0.002490i$ & $\;\;\;0.006990 + 0.002490i$ & $-9.537232 - 9.735256i$ \\
3 & 3 & 1 & 0.7 & $-0.088686 + 0.005499i$ & $-0.088686 + 0.005499i$ & $\;\;\;3.791563 - 2.194855i$ \\
 & & & 0.9999 & $-0.000594 - 0.047323i$ & $-0.000594 - 0.047323i$ & $\;\;\;0.000361 - 0.006627i$ \\
\end{tabular}
\end{ruledtabular}
\end{table*}

However, we both disagree with earlier literature such as Refs.~\cite{Glampedakis:2003dn, Zhang:2013ksa}.
Other than the fact that Ref.~\cite{Glampedakis:2003dn} did not make the distinction between Teukolsky \glspl{QNEF} and \gls{SN} \glspl{QNEF}, our discrepancies can be traced back to an ambiguity in the definition of the tortoise coordinate where there is a freedom to choose the integration constant while solving Eq.~\eqref{eq:drdrstar}.
The community has canonically chosen a particular integration constant such that $r_{*}(r)$ is given by Eq.~\eqref{eq:rstar_from_r}.
Note that this degree of freedom is equivalent to the one that allows us to choose the matching point $r_*^{\rm{mp}}$ [cf. Eq.~\eqref{eq:contour_rotation}] where choosing $r_*^{\rm{mp}} = 0$ is identical to choosing the integration constant like Eq.~\eqref{eq:rstar_from_r}.
Explicitly, if we denote a \gls{QNEF} computed with $r_*^{\rm{mp}} = k$ as $\left.B_{n}\right|_{r_*^{\rm{mp}} =\,k}$, then
\begin{equation}
\label{eq:QNEF_ambiguity}
	\left.B_{n}\right|_{r_*^{\rm{mp}} =\,0} = \left.B_{n}\right|_{r_*^{\rm{mp}} =\,k} e^{-2ik\omega_{n}}.
\end{equation}
This degeneracy was also noted in Ref.~\cite{Oshita:2024wgt} in a different context.
It is, therefore, essential to keep track of the choice(s)\footnote{This ``constant'' can be chosen differently depending on the value of $\omega_n$ so long as it is accounted for. In fact, we leverage this freedom to gain efficiency in computations (see the \hyperref[app:choosing_mp]{Appendix}).} made for this constant shift in the tortoise coordinate.

In Table \ref{tab:gravitational_QNEF}, again as an example, we show some of the values for both $B^{\text{T}, (+2)}_{n}$, which are for the first time computed explicitly, and $B^{\text{T}, (-2)}_{n}$. They are in excellent agreement with Eq.~\eqref{eq:Bpm2conv}. We have also compared our values, though not shown here, from our explicit calculations for $B^{\text{T},(+1)}_{n}$ with that converted from $B^{\text{T},(-1)}_{n}$ using Eq.~\eqref{eq:Bpm1conv} and also find excellent agreement.

\section{\label{sec:Discussions}Discussion of results}

\subsection{Behavior of \acrlong{QNM} trajectories in the complex plane}
From Figs.~\ref{fig:scalar_qnm}--\ref{fig:gravitational_qnm}, we can look for patterns in scalar, electromagnetic, and gravitational \gls{QNM} frequencies, respectively. 
We see that, for most of the $m \geq 0$ ``ordinary'' \glspl{QNM} (solid lines), their real part of the frequencies increases monotonically from $a = 0$ to their maximal values as $a \to 1$, while their imaginary part of the frequencies becomes less negative (i.e., longer damping times $\tau \equiv -1/\mathrm{Im}\;\omega$), also monotonically.
There are some notable exceptions to this, which will be discussed later in Sec.~\ref{subsubsec:crossing}.
For $m < 0$ ordinary modes, the trajectories are tangent to their positive $m$ counterparts at $a = 0$.
Their trends with increasing $a$ are, however, less obvious.
Usually for $m = -\ell$ modes, the real part of the frequencies becomes smaller and the imaginary part of the frequencies becomes more negative (i.e., shorter damping times) with increasing $a$, while for modes with lower $|m|$, the trends are completely opposite.

\subsubsection{\Acrlong{QNM} trajectories when approaching the extremal Kerr limit}
Most of the \glspl{QNM} with $\ell = |m| (\neq 0)$\footnote{Recall that there is another set of \gls{QNM} frequencies (mirror modes) with $\left\{ \omega^{-}_{s \ell m n} \equiv -{\omega_{s \ell (-m) n}^{+}}^{*} \right\}$ that are not shown explicitly in Figs.~\ref{fig:scalar_qnm}--\ref{fig:gravitational_qnm}.} tend to accumulate at $\omega_n = m/2$ when $a \to 1$, again, with some exceptions (e.g., $|s| = \ell = m = 2, n = 5$).
This is, however, not limited only to $\ell = |m|$ modes. For example, it can also be seen in the $|s| = 2, \ell = 7, m = 6$ trajectory in Fig.~\ref{fig:gravitational_qnm}.
As noted in Refs.~\cite{Yang:2012pj, daCunha:2021jkm} and clearly visible in Figs.~\ref{fig:scalar_qnm}--\ref{fig:gravitational_qnm}, there are two limiting behaviors or branches for the \glspl{QNM} when $a \to 1$, namely, zero-damping modes, where $\mathrm{Im}\;\omega_n \to 0$ and, thus, infinite damping times, and damped modes, where $\mathrm{Im}\;\omega_n\;\slashed{\to}\;0$.

While very tempting, we are not able to distinguish if a particular \gls{QNM} is a zero-damping one or a damped one, since our calculations are only up to $a = 1 - 10^{-4}$ and not exactly at $a = 1$. However, given that $\omega_n$ should be a smooth function of $a$, the \gls{QNM} frequency computed sufficiently close to $a = 1$ should be indicative \cite{Richartz:2015saa}. Therefore, if the limiting value for the imaginary part of a \gls{QNM} frequency is clearly away from 0 (e.g., the $s = \ell = m = 0, n = 1$ mode as shown in Fig.~\ref{fig:s0l0m0n1}), we can still confirm that a \gls{QNM} is a damped mode with our results.

\subsubsection{\label{subsubsec:crossing}Crossings of \acrlong{QNM} trajectories}
Some of the \gls{QNM} trajectories, however, show rather peculiar features as shown in Figs.~\ref{fig:scalar_qnm}--\ref{fig:gravitational_qnm} and deviate from the general trends described earlier in this section. For instance, the $s = \ell = m = 0, n = 1$ \gls{QNM} trajectory in Fig.~\ref{fig:s0l0m0n1} exhibits a spiraling motion in the complex frequency plane as $a$ approaches extremality forming multiple knots before settling down to the $a = 1$ limit, away from the real axis. This example is by no means isolated. In fact, there are many more modes (that are tabulated in this work) that show similar behaviors (e.g., $|s| = 2, \ell = 2, m = 0, n = 5, 6, 7$, to just name a few).

Another notable example includes the $|s| = \ell = m = 2, n = 5$ mode, which also forms a knot in the complex frequency plane and converges to a frequency with nonzero imaginary part at $a = 1$.
This particular mode received considerably more interest (e.g., Refs.~\cite{Onozawa:1996ux, Oshita:2021iyn, Motohashi:2024fwt}) because of its observational relevance by \gls{GW} detectors. 
It was recently remarked in a footnote in Ref.~\cite{Dias:2021yju} and further explored by Ref.~\cite{Motohashi:2024fwt} that the phenomenon of eigenvalue repulsion can explain this peculiarity.

Avoided crossing is well understood in the context of quantum mechanics (see, for example, Refs.~\cite{Landau:1991wop, Moiseyev_2011}). The real eigenvalues (corresponding to physical observables) of a quantum system described by a Hermitian Hamiltonian with $k$-many real parameters avoid crossing and repel from each other, except when the system has a degeneracy or on a $(k-2)$-dimensional manifold \cite{Landau:1991wop}.
In literature, this phenomenon goes by different names, including eigenvalue repulsion, level crossing, etc., that carry the same meaning in the Hermitian case.

For non-Hermitian systems, however, their eigenvalues can be complex.
In particular, the real part of a complex eigenvalue is referred to as the (energy) level, while the imaginary part is referred to as the width since the corresponding eigenfunction decays spatially.
These systems also exhibit eigenvalue repulsion.
However, the real part (i.e., the levels) or the imaginary part (i.e., the widths) of the eigenvalues can cross each other but \emph{not simultaneously}.
Unique to non-Hermitian systems is that, when two levels cross each other, the corresponding widths repel, and vice versa \cite{PhysRevE.61.929}.

Let us consider a simple two-level system where the (effective) Hamiltonian $H_{\text{eff}}$ is given by \cite{PhysRevE.61.929, Heiss:2012dx}
\begin{equation}
\label{eq:effective_Hamiltonian}
H_{\text{eff}} = 
\begin{pmatrix}
\epsilon_{1} & 0 \\
0 & \epsilon_{2}	
\end{pmatrix} + \lambda
\begin{pmatrix}
 V_{11} & V_{12} \\
 V_{21} & V_{22}
\end{pmatrix},
\end{equation}
where $V_{ij}, \epsilon_{1, 2} \in \mathbb{C}$ are constants and $\lambda \in \mathbb{C}$ is a ``control parameter'' that, in general, can be complex, though it is usually restricted to be real for physical systems.
The eigenvalues $E_{1,2}$ under this model is, thus, given by
\begin{multline}
\label{eq:Heiss_eigenvalues}
E_{1,2} = \dfrac{1}{2} \left. \biggl\{ \lambda \left( V_{11} + V_{22} \right) + \left( \epsilon_{1} + \epsilon_{2} \right) \right. \\
\left. \mp \sqrt{4\lambda^2 V_{12}V_{21} + \left[ \lambda\left(V_{11} - V_{22} \right) + \left( \epsilon_1 - \epsilon_2 \right) \right]^2} \right. \biggl\}.
\end{multline}
When $\lambda$ is allowed to be complex, the Hamiltonian can be thought of as having two real parameters $\mathrm{Re}\;\lambda$ and $\mathrm{Im}\;\lambda$, respectively.
This means that the two eigenvalues can now cross on a $0$-dimensional manifold (i.e., discrete disconnected points), referred to as exceptional points in the literature \cite{kato1995perturbation, Heiss:2012dx}.
For this two-level system, the eigenvalues cross when the expression inside the square root in Eq.~\eqref{eq:Heiss_eigenvalues} vanishes. This means that the exceptional points $\lambda_{\text{EP}}$ of this model are located at
\begin{equation}
\label{eq:exceptional_points}
	\lambda_{\text{EP}} = \dfrac{-\left( \epsilon_1 - \epsilon_2 \right)\left(V_{11} - V_{22}\right) \pm 2i \left( \epsilon_1 - \epsilon_2 \right) \sqrt{V_{12}V_{21}} }{4V_{12}V_{21} + \left(V_{11} - V_{22}\right)^2}.
\end{equation}

\begin{figure*}
\centering
\includegraphics[scale=0.5]{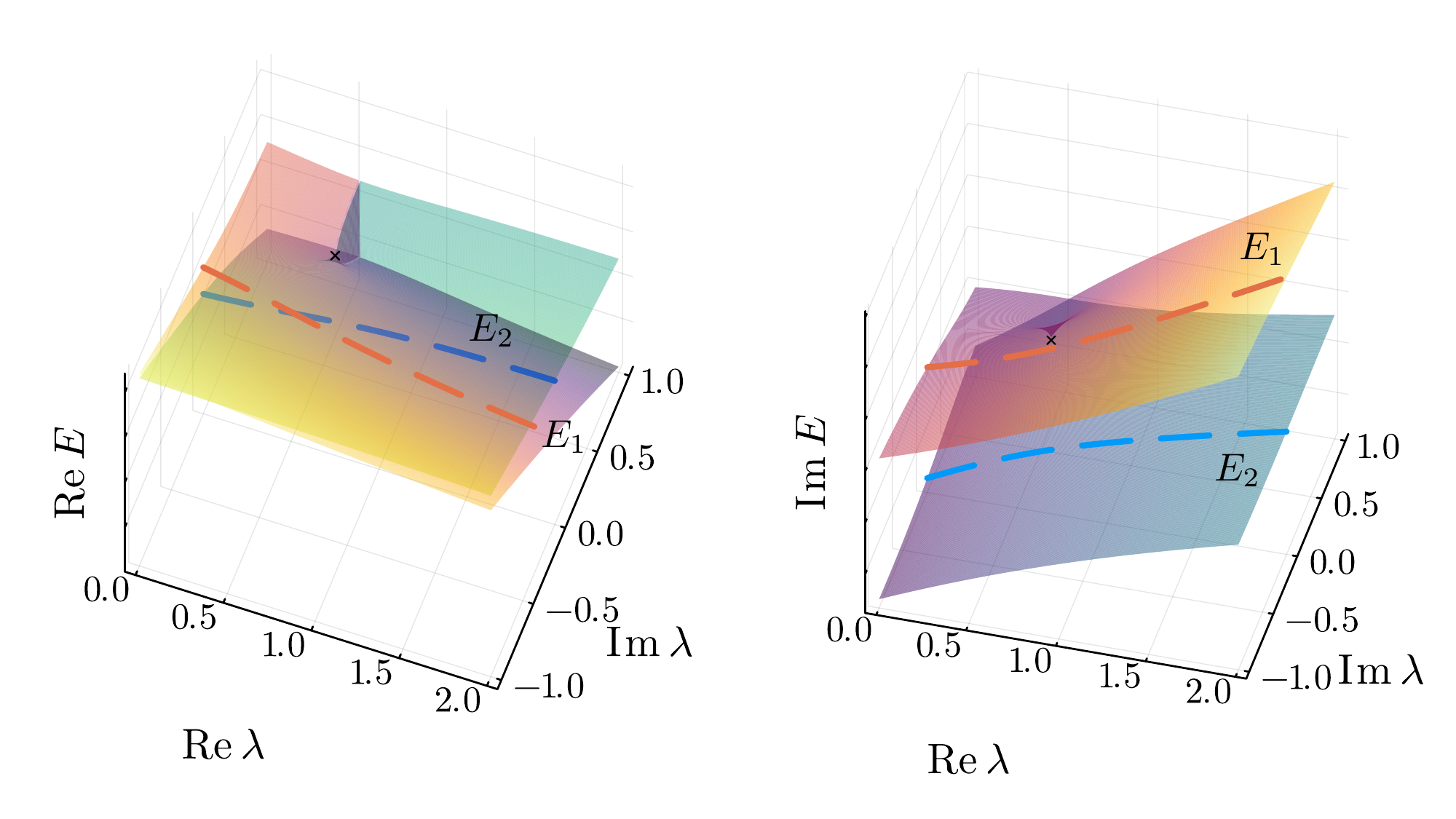}
\caption{\label{fig:Riemann_surface}Two eigenvalues $E_{1,2}(\lambda \in \mathbb{C})$ visualized as two Riemann surfaces that intersect at exceptional points (left panel, the real part; right panel, the imaginary part). In this example, one of the exceptional points is located at around $\lambda_{\text{EP}} \approx 0.52 + 0.62i$, which is indicated by a cross. The dashed lines correspond to the two physically relevant slices $E_{1,2}(\lambda \in \mathbb{R})$ along $\mathrm{Im}\;\lambda = 0$.}	
\end{figure*}

In fact, the two eigenvalues $E_{1,2}(\lambda \in \mathbb{C})$ can be visualized as two Riemann surfaces that intersect at $\lambda = \lambda_{\text{EP}}$.
Importantly, these Riemann surfaces \emph{smoothly} connect different branches of the multivalued square root function in Eq.~\eqref{eq:Heiss_eigenvalues} across its branch points, which exactly correspond to the exceptional points.
This is illustrated in Fig.~\ref{fig:Riemann_surface}, where we plot $\mathrm{Re}\;E_{1,2}(\lambda \in \mathbb{C})$ (left panel) and $\mathrm{Im}\;E_{1,2}(\lambda \in \mathbb{C})$ (right panel) as two surfaces, separately and respectively.
In this example, one of the branch points or exceptional points is located at around $\lambda_{\text{EP}} \approx 0.52 + 0.62i$ (indicated by a cross).
Following the upper surface in the right panel in Fig.~\ref{fig:Riemann_surface} that corresponds to $E_{1}$ from the left to the right across the branch point, one needs to choose the branch that has an opposite sign to the branch chosen left to the branch point to ensure continuity (and vice versa for the lower surface for $E_2$).
This is the mathematical origin of eigenvalue repulsion, stemming from the continuity of the eigenvalues in the control parameter $\lambda$.
The real part of the eigenvalues $\mathrm{Re}\;E_{1,2}$ as shown in the left panel in Fig.~\ref{fig:Riemann_surface} cross each other around the branch point with the properly chosen branches.
It had also been demonstrated that the same intuition holds even when there are more than two levels in a system \cite{PhysRevE.61.929}.

In the context of \glspl{BH}, the linear differential operators associated with their (linear) perturbations are non-Hermitian, also evident from the fact that their \gls{QNM} frequencies are complex.
Eigenvalue repulsion has been reported in the \gls{QNM} spectra of charged rotating \glspl{BH} \cite{Dias:2021yju}, as well as rotating \glspl{BH} coupled with a massive scalar field of mass $\mu$ \cite{Cavalcante:2024swt}.
In both cases, the spectra are parametrized by two real parameters.
Moreover, in the later case, an exceptional point located at real values of $a$ and $\mu$ was found for the first time \cite{Cavalcante:2024swt}.
The reality of the exceptional point has implications beyond simply confirming the mathematical model of avoided crossing. Physically, this leads to hysteresis where the state of a \gls{BH} depends on the path that it has taken around such an exceptional point, which can be realized by a series of physical processes such as varying the angular momentum of the \gls{BH} and the mass of the scalar field.

From our discussions, we can see that eigenvalue repulsion is a generic feature ubiquitous in eigensystems. It originates from branch point singularities and has little to do with the nature of gravity.
Here, we analyze two examples of the crossing of the real or the imaginary part of the \gls{QNM} trajectories (but not simultaneously) using the mathematical theory of eigenvalue repulsion.

\begin{figure*}
\subfloat[\label{fig:two_level_model_fit_real}]{\includegraphics[width=\columnwidth]{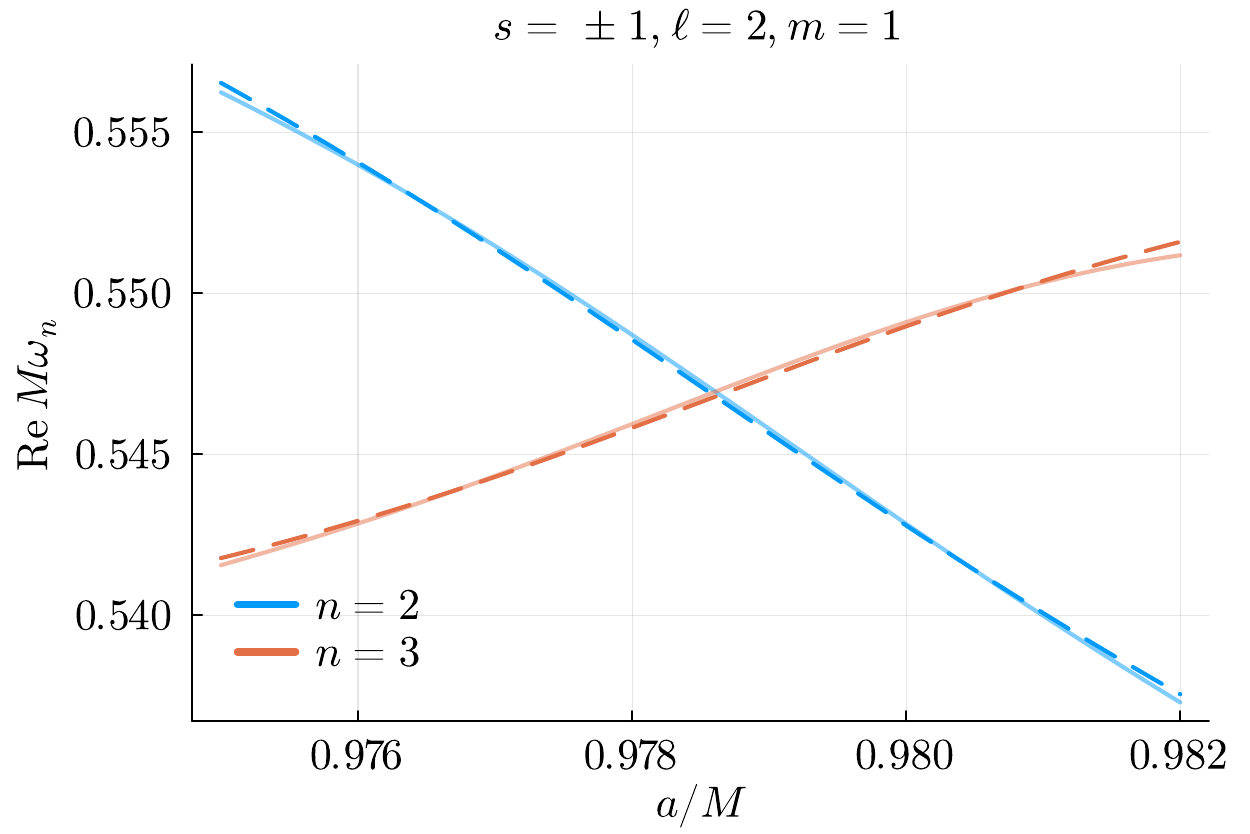}}
\subfloat[\label{fig:two_level_model_fit_imag}]{\includegraphics[width=\columnwidth]{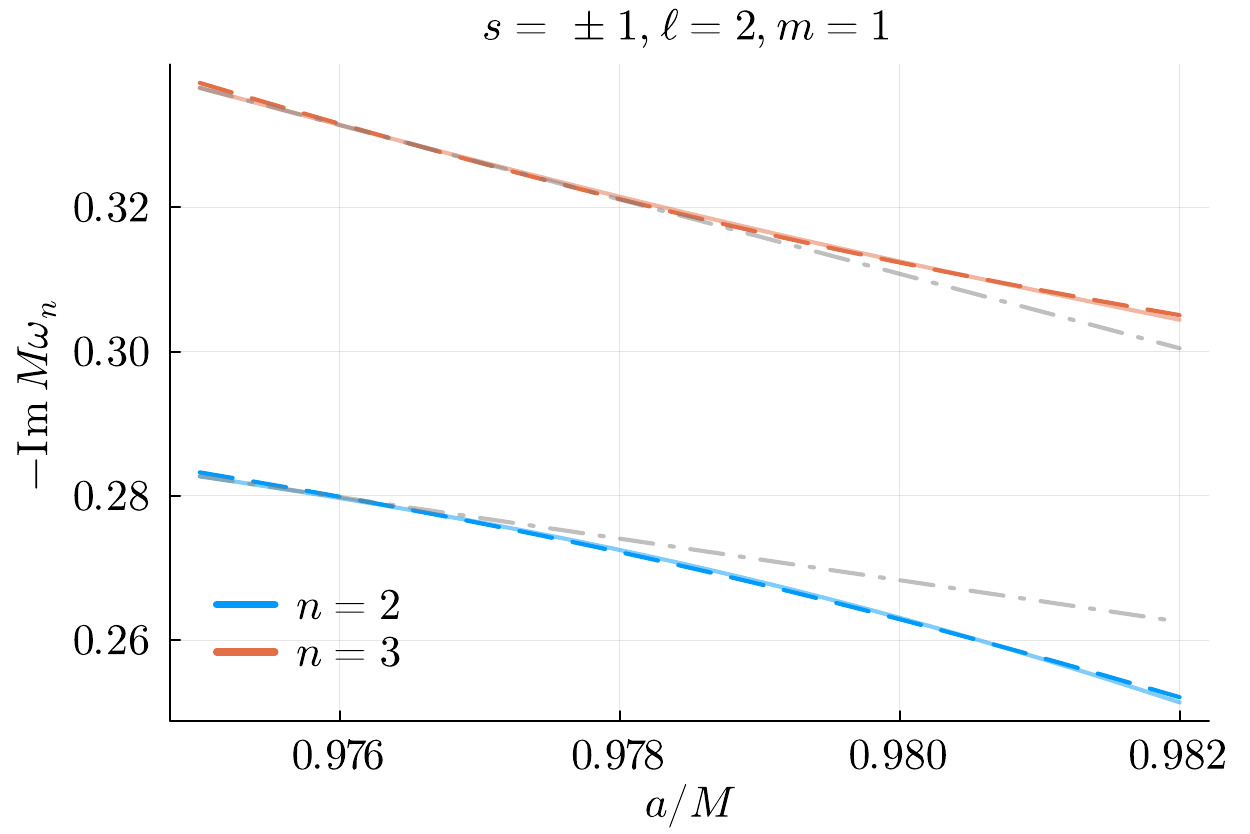}}
\caption{\label{fig:two_level_model_fit}Real part (a) and imaginary part (b) of the \gls{QNM} frequencies of the $s = \pm 1, \ell = 2, m = 1$ with $n = 5$ and $n = 6$ from $a/M = 0.975$ to $a/M = 0.982$, respectively. The solid curves in both panels show the actual values for the frequencies, while the dashed curves show the best-fit results using the two-level effective Hamiltonian in Eq.~\eqref{eq:effective_Hamiltonian}. With only ten degrees of freedom, the model reproduces the crossing in the real part of the frequencies and the corresponding repulsion in the imaginary part of the frequencies. The dash-dotted lines in (b) are the tangents of the curves at $a/M = 0.975$ to contrast the repulsion among the two overtones more clearly.}
\end{figure*}

The first example that we consider is the $s = \pm 1, \ell = 2, m = 1$ mode. The real part of the \gls{QNM} frequency of the $n = 2, 3$ overtones cross each other at around $a \approx 0.9785$ (see the blue and orange solid curves in Fig.~\ref{fig:two_level_model_fit_real}, respectively). The imaginary part of the frequencies can be seen to repel each other slightly at around the same value of $a$ in Fig.~\ref{fig:two_level_model_fit_imag}.
As a first attempt, we fit the complex \gls{QNM} frequencies \emph{locally} around this crossing point (from $a_{0}=0.975$ to $a_{1}=0.982$, for a total of $142$ complex frequencies) with the two-level effective Hamiltonian in Eq.~\eqref{eq:effective_Hamiltonian} where its eigenvalues $E_{1,2}$ model the complex \gls{QNM} frequencies, i.e., $E_{1,2} \equiv \omega_{1,2}$.\footnote{One could have also chosen the eigenvalues as $E_{1,2} = \left(\omega_{1,2}\right)^2$ instead, as one can recast the \gls{GSN} equation into a form that resembles the Schrödinger equation, i.e., $d^2 Y/dr_{*}^2 + \left[\omega^2 - V_{Y}(r_*)\right]Y = 0$, where $Y \equiv X/\sqrt{\eta}$ (note that $\eta$ is frequency dependent) and $V_Y$ can be expressed in terms of $\mathcal{F}$ and $\mathcal{U}$ \cite{Lo:2023fvv}. Since we do not attempt to map the two-level model back into the \gls{GSN} equation, both choices are valid, and we chose $E_{1,2} = \omega_{1,2}$ for ease of demonstration.}
The control parameter $\lambda$ in this case is simply $a$ rescaled such that it is bounded between $\lambda \in [0, 1]$, i.e., $\lambda \equiv (a - a_0)/(a_1 - a_0)$.
Note that there are only ten real variables to be fitted (instead of $12$) in this phenomenological model. This is because, as far as eigenvalues are concerned, $V_{12}$ and $V_{21}$ enter only as the product $V_{12}V_{21}$ in Eq.~\eqref{eq:Heiss_eigenvalues}. In fact, we will set $V_{12} = V_{21}$ without loss of generality hereinafter.

The best-fit results for the two overtones are shown as the dashed curves in Figs.~\ref{fig:two_level_model_fit_real} and~\ref{fig:two_level_model_fit_imag}, respectively.
We see that the phenomenological model fits the \gls{QNM} frequency data very well, reproducing the crossing in the real part of the frequencies and the repulsion in the imaginary part of the frequencies with just ten degrees of freedom.\footnote{One could even argue that there are effectively only six degrees of freedom instead, since $\epsilon_{1,2} \in \mathbb{C}$ can be read off or determined simply from the $a = a_0$ values.}
Note that the model needs to fit simultaneously the real and the imaginary parts of both \glspl{QNM}. Therefore, the excellent agreement of the fit with the data is unlike to be due to overfitting.
From the model, we can also infer that there are two exceptional points in the complex-$a$ parameter space at roughly $a_{\text{EP}}^{(1)} \approx 0.979 - 0.024i$ with $\omega_{\text{EP}}^{(1)} \approx 0.654 - 0.275i$ and $a_{\text{EP}}^{(2)} \approx 0.979 + 0.004i$ with $\omega_{\text{EP}}^{(2)} \approx 0.527 - 0.297i$, respectively. Unlike Ref.~\cite{Cavalcante:2024swt}, these exceptional points \emph{cannot} be encircled via physical processes (i.e., spinning a \gls{BH} up or down) and, thus, are not physically interesting.
That said, their mathematical existence still influences the \gls{QNM} trajectories as shown here and can be confirmed by analytically continuing the Teukolsky solutions for a complex value of $a$. We leave this for future work.

\begin{figure*}
\subfloat[\label{fig:two_level_model_fit_real_grav}]{\includegraphics[width=\columnwidth]{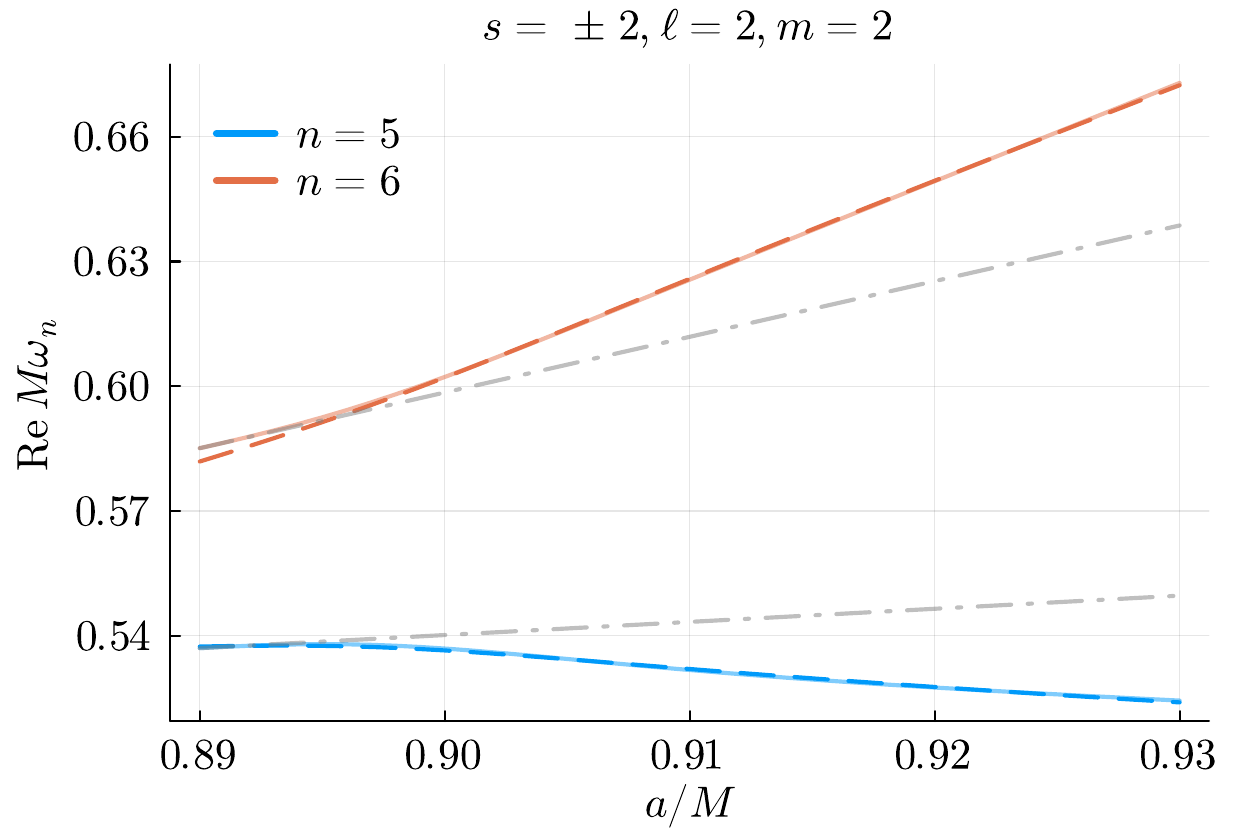}}
\subfloat[\label{fig:two_level_model_fit_imag_grav}]{\includegraphics[width=\columnwidth]{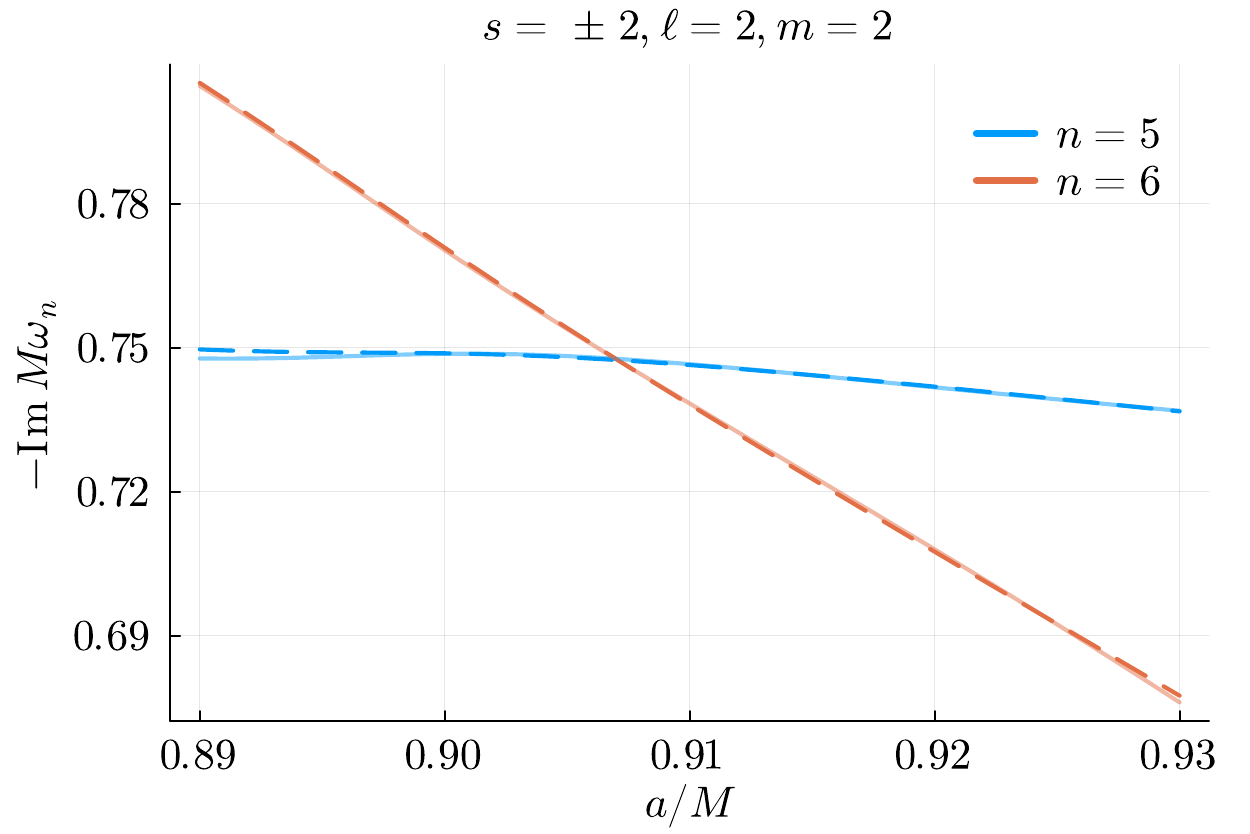}}
\caption{The same as Fig.~\ref{fig:two_level_model_fit} with (a) showing the real part of the frequencies and (b) showing the imaginary part of the frequencies, respectively, but for $|s| = \ell = m = 2$ and $n = 5, 6$ \glspl{QNM}.}
\end{figure*}
To demonstrate that the agreement is not just accidental, we repeat the exercise with another example, the $|s| = \ell = m = 2$ and $n = 5, 6$ modes.
This time we fit the \glspl{QNM} with the two-level effective Hamiltonian using even more data from $a_0 = 0.89$ to $a_1 = 0.93$ for a total of $622$ complex frequencies.
The best-fit results are shown as the dashed curves in Figs.~\ref{fig:two_level_model_fit_real_grav} and~\ref{fig:two_level_model_fit_imag_grav}, respectively.
Once again, the model fits the data very well (solid curves in both figures) and reproduces the crossing in the imaginary part of the frequencies and the corresponding repulsion in the real part of the frequencies with only a few degrees of freedom.
Using the model, we infer that there are two exceptional points in the complex-$a$ parameter space at roughly $a_{\text{EP}}^{(1)} \approx 0.897 + 0.010i$ with $\omega_{\text{EP}}^{(1)} \approx 0.550 - 0.755i$ and $a_{\text{EP}}^{(2)} \approx 0.876 + 0.032i$ with $\omega_{\text{EP}}^{(2)} \approx 0.490 - 0.772i$, respectively.
Despite being able to easily explain the \gls{QNM} trajectories near the point where the real and imaginary parts of the frequencies cross, we want to caution with the caveats that the two-level model does not take into account interactions from other overtones (in fact, the $n = 5$ overtone crosses both $n = 6$ and $n = 7$ overtones but at different values of $a$) and that the model is only a local expansion of the much more complicated eigensystem.

\subsubsection{Resonant excitation of \acrlongpl{QNM}}

Most of the \gls{QNEF} trajectories, parametrized by $a$, computed in this work are either in a spiral motion (also noted in Ref.~\cite{Motohashi:2024fwt}) or in a parabolic/hyperboliclike motion in the complex plane as $a$ increases from $0$ to $0.9999$.
However, like the \gls{QNM} trajectories, some of the \gls{QNEF} trajectories show anomalous features.

\begin{figure*}
\includegraphics[width=2\columnwidth]{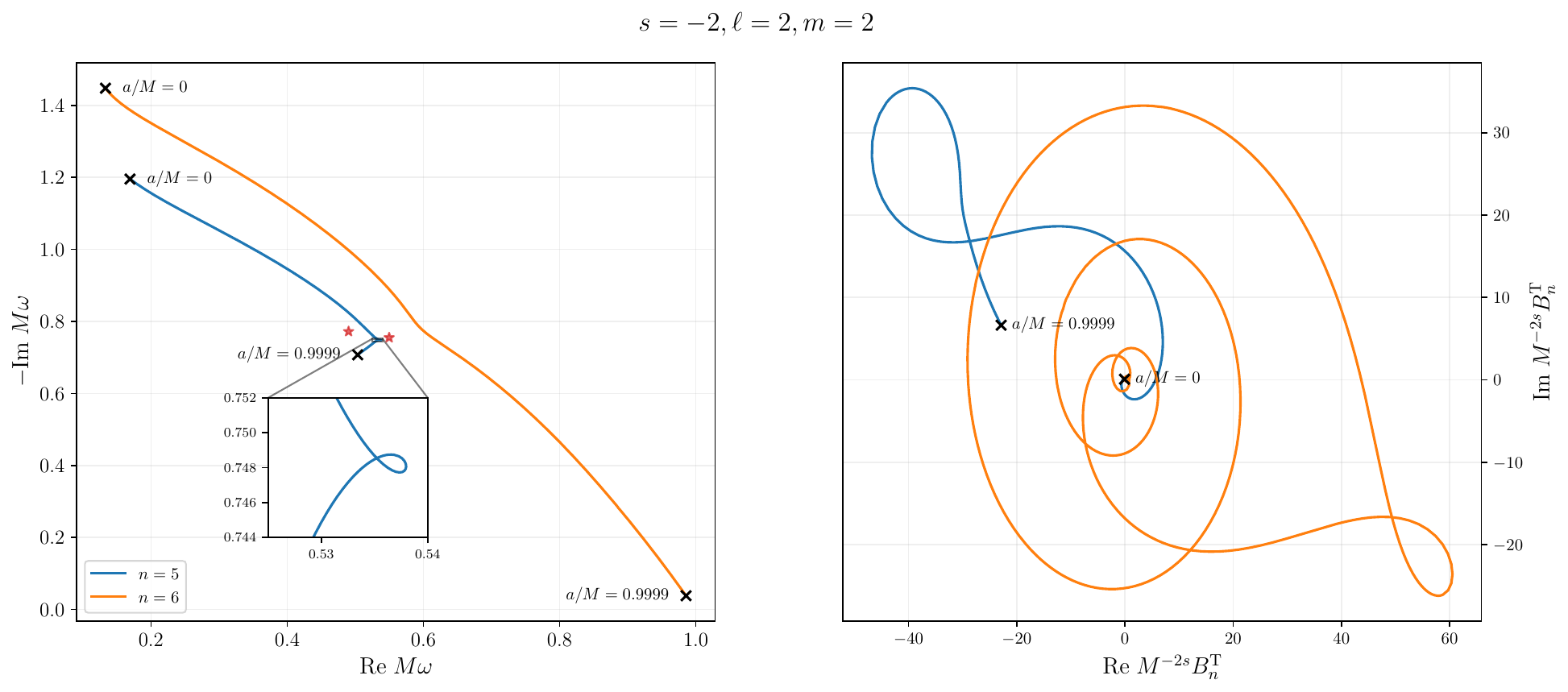}
\caption{\label{fig:gravitational_crossing_with_resonance}\gls{QNM} (left panel) and \gls{QNEF} (right panel) trajectories, parametrized by $a/M$, for the $s = -2, \ell = m = 2$ and $n = 5, 6$ modes, respectively. The two frequencies $\omega_{\text{EP}}$ corresponding to the exceptional points inferred using the two-level model in Sec.~\ref{subsubsec:crossing} are shown by stars in the left panel.}
\end{figure*}

For instance, Fig.~\ref{fig:gravitational_crossing_with_resonance} shows the \gls{QNM} (left panel) and the \gls{QNEF} (right panel) trajectories for the $s = -2, \ell = m = 2$ and $n = 5, 6$ modes.
The two \gls{QNEF} trajectories, as first pointed out in Ref.~\cite{Motohashi:2024fwt}, spiral against each other in a lemniscatelike pattern (also known as a ``figure-8'' pattern) near the value of $a \approx 0.897$ where the $n = 5$ \gls{QNM} trajectory forms a knot. As shown explicitly in Sec.~\ref{subsubsec:crossing}, this is related to eigenvalue repulsion near exceptional points.

\begin{figure*}
\centering
\includegraphics[width=2\columnwidth]{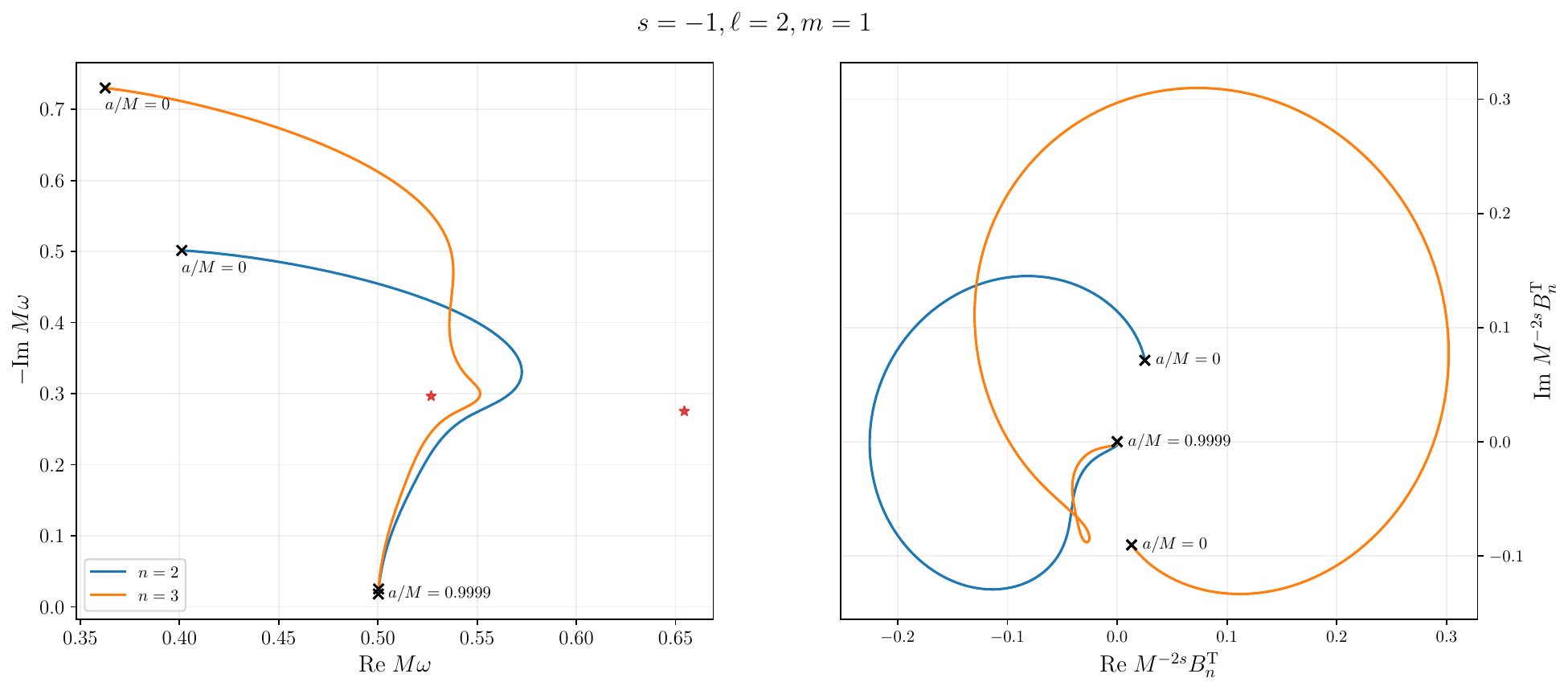}
\caption{\label{fig:electromagnetic_crossing_with_resonance}The same as Fig.~\ref{fig:gravitational_crossing_with_resonance} but with the $s = -1, \ell = 2, m = 1$ and $n = 2, 3$ modes, respectively. However, we do not see a complete lemniscate in the \gls{QNEF} trajectories as proposed by Ref.~\cite{Motohashi:2024fwt} that signifies the occurrence of a resonant excitation. This is likely due to the greater distance of the \gls{QNM} trajectories in the left panel with the frequencies corresponding to the exceptional points as shown by the stars compared to the gravitational case in Fig.~\ref{fig:gravitational_crossing_with_resonance}.}
\end{figure*}

Figure~\ref{fig:electromagnetic_crossing_with_resonance} shows another example for electromagnetic \glspl{QNM} with $s = -1, \ell = 2, m = 1$ and $n = 2, 3$.
Similar to Fig.~\ref{fig:gravitational_crossing_with_resonance}, these two modes show eigenvalue repulsion as demonstrated in Sec.~\ref{subsubsec:crossing}. However, unlike Fig.~\ref{fig:gravitational_crossing_with_resonance}, we do not see a complete lemniscate in the \gls{QNEF} trajectories that Ref.~\cite{Motohashi:2024fwt} suggests to signify the occurrence of a resonant excitation near exceptional points.
This is likely due to the fact that the two exceptional points as inferred by the two-level effective Hamiltonian model are further away from the \gls{QNM} trajectories compared to the gravitational case shown in Fig.~\ref{fig:gravitational_crossing_with_resonance}, and, thus, are amplified to a lesser extent.


\subsection{Comparisons with other methods}
Besides the method presented here, there are other techniques for computing \gls{QNM} frequencies and \glspl{QNEF} of Kerr \glspl{BH}, each with their strengths and weaknesses.
Here, we outline a few of those and compare them with our method.

Continued-fraction-based methods (e.g., Refs.~\cite{Leaver:1985ax, Cook:2014cta}) write solutions to the Teukolsky equation using a (truncated) finite number of basis functions.
The boundary conditions for \glspl{QNM} are imposed directly to find a recurrence relation, which is solved efficiently by transforming it into a continued fraction equation.
As a result, these methods converge very quickly and can produce high-precision \gls{QNM} frequencies (for one such numerical implementation, see Ref.~\cite{Stein:2019mop}).
However, they are incapable of computing \glspl{QNEF}.

\gls{MST}-based methods (e.g.~Refs.~\cite{Zhang:2013ksa, Casals:2019vdb, Motohashi:2024fwt}), on the other hand, leverage the analytic expressions for $B^{\text{inc, ref}}(\omega)$ [or, equivalently, $A_{\text{in, out}}(\omega)$] that one can easily evaluate given the values of $s, \ell, m$, and $a$, respectively.
They are very similar to our method, as \gls{QNM} frequencies are also found by numerically solving a nonlinear equation $A_{\text{in}}(\omega) = 0$. However, in our method, the left-hand side of the nonlinear equation, $A_{\text{in}}(\omega)$, itself is obtained by numerically integrating the \gls{GSN} equation. 
These techniques can compute \glspl{QNEF} quickly, since the asymptotic amplitudes are given as analytical expressions in terms of $\omega$.
However, the \gls{MST} method is a highly specialized and specific technique devised for Kerr \glspl{BH}, and extending the method to other backgrounds might be nontrivial.

Recently, isomonodromic and gauge-theory-inspired methods (e.g., Refs.~\cite{Aminov:2020yma, Bonelli:2021uvf, daCunha:2021jkm}) have been proposed.
They provide an exact quantization rule for \glspl{QNM} that one can solve directly for a given overtone number $n$ (though that might not match our convention for labeling overtones) instead of searching for it.
Exact expressions for the asymptotic amplitudes are also provided from these techniques.
More importantly, they give us a better analytical understanding of the behaviors of \glspl{QNM} and their \glspl{QNEF} than numerical explorations like this paper can offer.

Another recent trend in the computational techniques for Kerr \glspl{BH} is the adaptation of hyperboloidal coordinates (e.g., \cite{PanossoMacedo:2019npm, Ripley:2022ypi}).
Under the hyperboloidal framework, coordinate transformations are performed to eliminate some of the coordinate artifacts associated with the conventional Boyer-Lindquist coordinates.
In particular, \gls{QNM} solutions are regular at the two boundaries.
Coupled with a (pseudo)spectral algorithm, one will also have to solve an eigensystem numerically to obtain \gls{QNM} frequencies. For one such numerical implementation, see Ref.~\cite{Ripley:2022ypi}.
Our method is conceptually similar where a coordinate transformation [i.e., a complex scaling transformation; cf. Eq.~\eqref{eq:contour_rotation}] is performed to get rid of unwanted coordinate artifacts near the boundaries. 

The complex scaling transformation is also employed in what their authors referred to as the ``quick-and-dirty'' method \cite{Glampedakis:2003dn, Silva:2024ffz}.
Indeed, our method bears many resemblances to it in many regards.
The key ingredient in such a scheme, as demonstrated in this paper, is really just the complex scaling transformation, and there is no need to use the Pr\"{u}fer phase transformation.
In addition, the matching point for the transformation needs to be chosen carefully and accounted for when comparing with other techniques of computing \glspl{QNEF} since their values depend on the definition of $r_{*}$ [cf. Eq.~\eqref{eq:QNEF_ambiguity}]. With these improvements, as well as the switch to solving the \gls{GSN} equation numerically instead of using the phase integral method, we show that one can reliably push the calculations all the way to near extremal rotation (in this work, to $a = 1 - 10^{-4}$ systematically) and moderate overtones (in this work, up to $n \leq 7$)\footnote{Although not shown in this paper and not done systematically, we did try computing overtones up to $n = 20$ and found no issue.} unlike what was claimed in Ref.~\cite{Glampedakis:2003dn}.

\section{\label{sec:conclusion}Conclusion}
In this work, we introduce a novel technique based on the \gls{GSN} formalism to compute the values of the frequencies and the corresponding \glspl{QNEF} for scalar, electromagnetic, and gravitational \glspl{QNM}.
Specifically, we analytic continue the tortoise coordinate so that \gls{GSN} solutions remain finite at the horizon and near spatial infinity.

Our results for the \gls{QNM} frequencies and their \glspl{QNEF}, in general, agree with previous calculations that were done with independent methods, thus cross-validating values found in the literature.
Furthermore, we clarify that \glspl{QNEF} also depend on the sign of the spin weight of the perturbing field $s$ and that the \glspl{QNEF} in the Teukolsky formalism are not dimensionless.

Using our data, we further explore the idea that the eigenvalue repulsion theory can explain some of the anomalous \gls{QNM} and \gls{QNEF} trajectories in the complex plane.
With a simple two-level effective Hamiltonian, we explicitly demonstrate, using two sets of \gls{QNM} trajectories as examples, that whenever the real parts of the eigenvalues cross, the imaginary parts of them repel but never simultaneously for real values of $a$, and vice versa.
Moreover, we show that the extent of the resonant excitation due to eigenvalue repulsion depends also on the proximity to the exceptional point that causes the repulsion.

The approach here can also be extended straightforwardly to compute \glspl{QNM} and \glspl{QNEF} of arbitrary backgrounds that admit a second-order \gls{ODE} as the master equation governing their perturbations.
Given such a master equation, apply an isospectral generalized Darboux transformation \cite{Glampedakis:2017rar} if necessary to ensure the potential associated with the \gls{ODE} is short-ranged. Then, use the complex scaling transformation to obtain a transformed \gls{ODE} that can be solved easily with conventional numerical methods.

\section*{Data Availability}
The data that support the findings of this article are openly available \cite{qnmqnefdata, qnminteractiveplot}.

\begin{acknowledgments}
The Center of Gravity is a Center of Excellence funded by the Danish National Research Foundation under Grant No. 184.
This work was supported by the research Grants No.~VIL37766 and No.~VIL53101 from Villum Fonden, and the DNRF Chair program Grant No. DNRF162 by the Danish National Research Foundation.
This work has received funding from the European Union's Horizon 2020 research and innovation program under the Marie Sklodowska-Curie Grant Agreement No.~101131233.
The Tycho supercomputer hosted at the SCIENCE HPC center at the University of Copenhagen was used for supporting this work.
Additionally, RKLL thanks Hayato Motohashi and Naritaka Oshita for their correspondence.
LS acknowledges support from the Stefan Rozental og Hanna Kobylinski Rozentals Fond.
\end{acknowledgments}

\appendix
\section*{\label{app:choosing_mp}Appendix: Choosing an appropriate matching point $r_{*}^{\text{mp}}$}
Choosing a different matching point $r_{*}^{\text{mp}}$ and rotation angle $\beta$ generates a different \emph{slice} of the analytic-continued $r(r_{*})$ as a function of a real variable $\rho$ [cf. Eq.~\eqref{eq:contour_rotation}].
In Sec.~\ref{subsec:GSN_analytic_continuation}, we give the appropriate value of $\beta$ to rotate such that a \gls{GSN} function $X(\rho)$ still behaves like a plane wave asymptotically (i.e., $\rho \to \pm \infty$) even with a complex frequency.
In this appendix, we focus on choosing the matching point $r_{*}^{\text{mp}}$ appropriately.

\begin{figure}[h]
\includegraphics[width=\columnwidth]{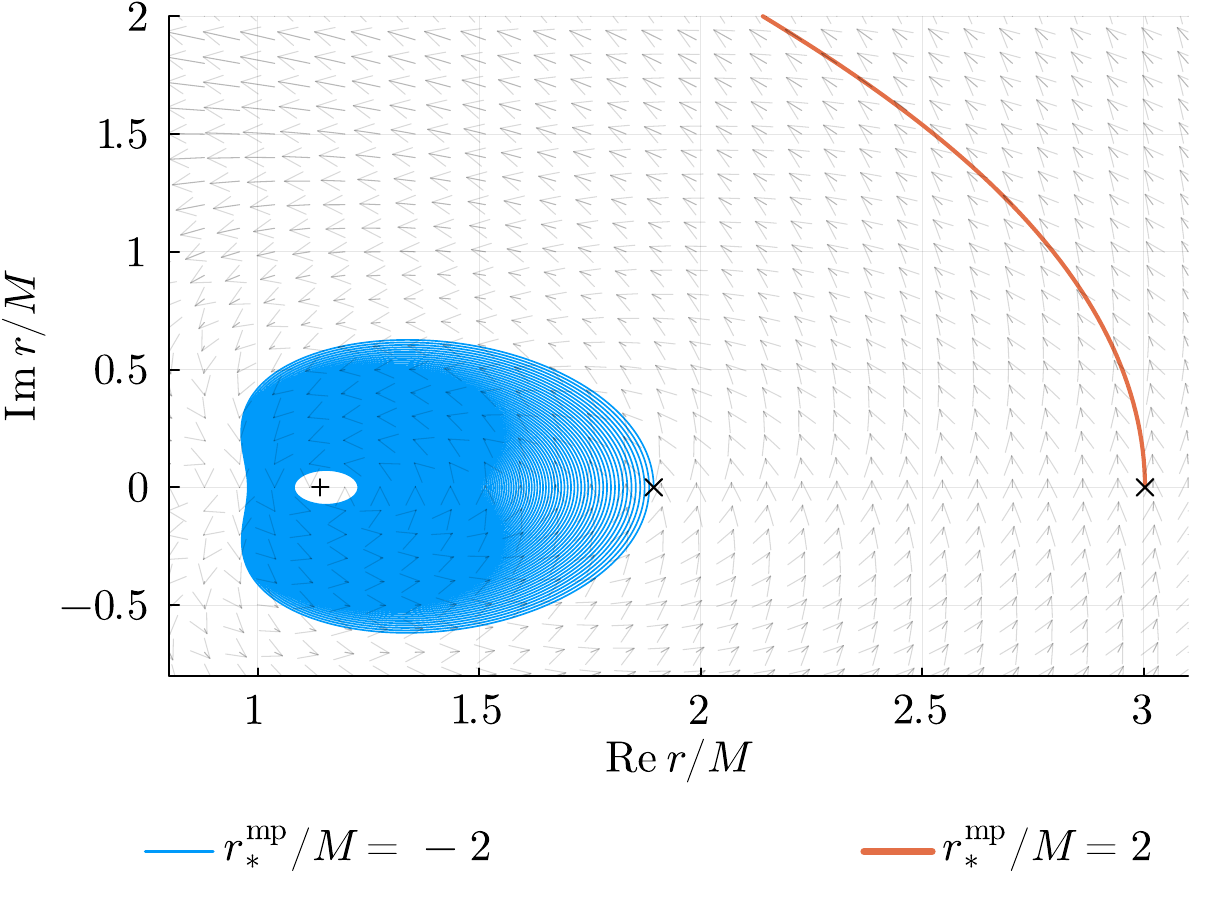}
\caption{\label{fig:r_from_rho_neg_different_rsmp}Two trajectories of $r(\rho)$ with matching point $r_{*}^{\text{mp}} = 2$ and $-2$, respectively, as $\rho$ changes from 0 (crosses) to be more negative.}
\end{figure}

A desirable $r(\rho)$ should map $\rho \to -\infty$ to $r \to r_{+} $ and $\rho \to +\infty$ to $|r| \to \infty$, respectively, such that Eq.~\eqref{eq:asymptotic_potentials} is satisfied.
In fact, only certain ranges of the matching point $r_{*}^{\text{mp}}$ give a $r(\rho)$ that has these properties.
Figure~\ref{fig:r_from_rho_neg_different_rsmp} shows $r(\rho \leq 0)$ for $a = 0.99$ and $\beta \approx -0.4996\pi$ (appropriate for $m = 2$ and $\omega \approx 0.8662 - 1.178i$) with two different matching points (indicated by the crosses at their corresponding values of $r^{\text{mp}}$ in the complex-$r$ plane), $r_{*}^{\text{mp}} = $ $2$ and $-2$, respectively, as an example.
We see that if we choose the matching point to be at $r_{*}^{\text{mp}} = 2$, then asymptotically $|r(\rho \to -\infty)| \to \infty$.
If we choose the matching point to be at $r_{*}^{\text{mp}} = -2$ instead, then $r$ approaches to $r_{+}$ (indicated by the plus sign) as desired.

These behaviors can be understood qualitatively by realizing that Eq.~\eqref{eq:drdrho} is an autonomous \gls{ODE}. The arrows in Fig.~\ref{fig:r_from_rho_neg_different_rsmp} indicate $-dr/d\rho$ at each point and the solutions $r(\rho)$ follow these arrows. There is a critical point on the real $r$ axis (which can be converted to a critical value of $r_{*}$) where all matching points to its left correspond to counterclockwise trajectories that asymptotically approach an equilibrium point on the real axis where $dr/d\rho = 0$, in this case at $r = r_{+}$.

\begin{figure}[h]
\includegraphics[width=\columnwidth]{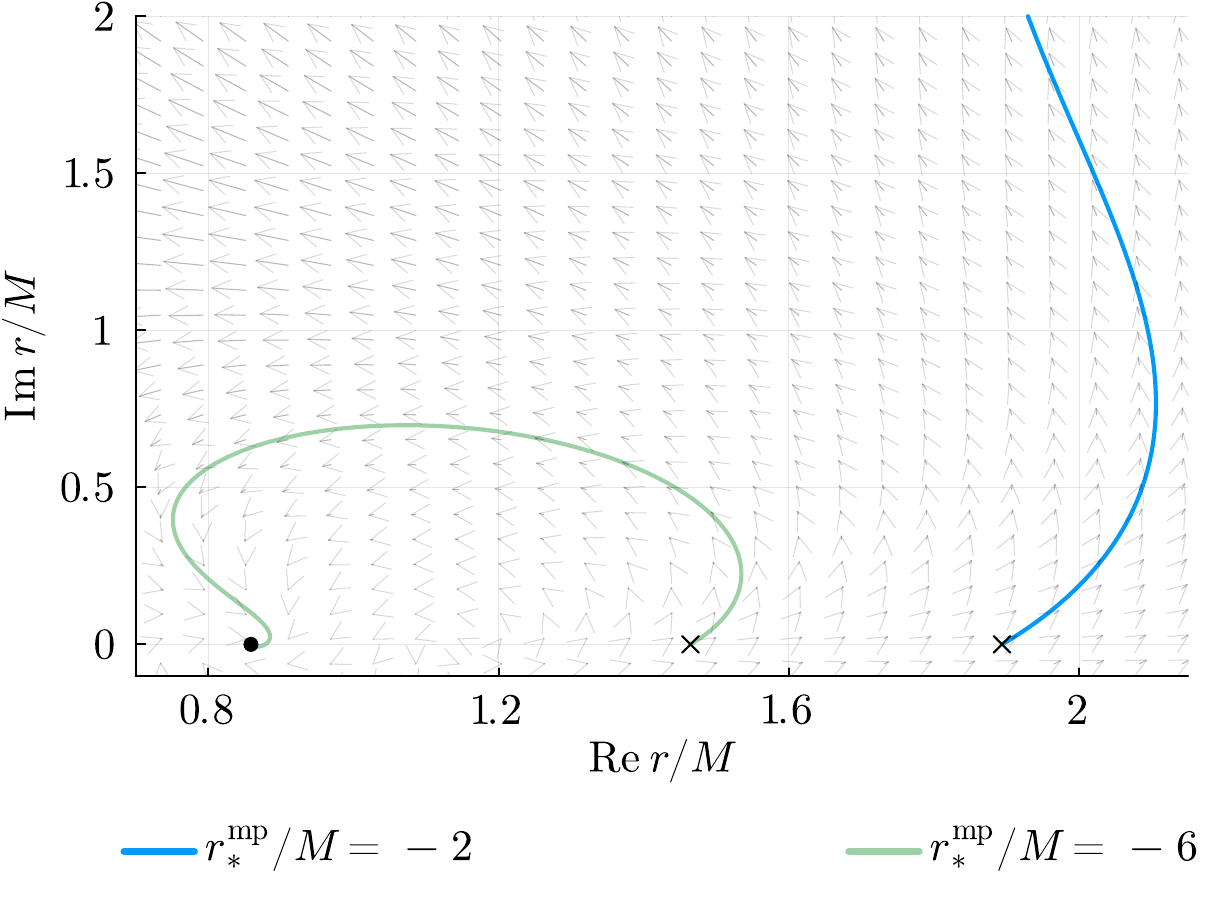}
\caption{\label{fig:r_from_rho_pos_different_rsmp}Two trajectories of $r(\rho)$ with matching point $r_{*}^{\text{mp}} = -2$ and $-6$, respectively, as $\rho$ changes from 0 (crosses) to be more positive.}
\end{figure}

One might then conclude that the matching point $r_{*}^{\text{mp}}$ can be chosen as small and far away from the critical value as one likes.
However, we need to consider also the asymptotic behavior of $r(\rho \to \infty)$.
Figure~\ref{fig:r_from_rho_pos_different_rsmp} shows explicitly two $r(\rho \geq 0)$ with $r_{*}^{\text{mp}} = $ $-2$ and $-6$, respectively, for $\beta \approx 0.2982\pi$, which is the appropriate value for the same $a$, $m$ and $\omega$ assumed in Fig.~\ref{fig:r_from_rho_neg_different_rsmp}.
The arrows in the figure indicate $dr/d\rho$ at each point such that the solutions $r(\rho \geq 0)$ follow these arrows.
We see that if the matching point (indicated by a cross) is not chosen correctly, say, at $r_{*}^{\text{mp}} = -6$ in this case, then $r(\rho)$ approaches to an equilibrium point on the real axis (in this case, at $r = r_{-}$ as indicated by a dot).
On the other hand, with $r_{*}^{\text{mp}} = -2$, $r(\rho \to \infty)$ has the desired trajectory asymptotically as $\rho \to \infty$.

Similar to the case with $\rho \leq 0$ as shown in Fig.~\ref{fig:r_from_rho_neg_different_rsmp}, there is also a critical point in Fig.~\ref{fig:r_from_rho_pos_different_rsmp} for the $\rho \geq 0$ case (which is different, since $\beta$ are not the same) where trajectories starting to the right of the critical point all approach to infinity asymptotically. Therefore, we can set the matching point $r_{*}^{\text{mp}}$ anywhere between those two critical points.

Ideally, one would want to choose $r_{*}^{\text{mp}}$ as the lower limit of the allowed range such that the corresponding $r^{\text{mp}}$ in the complex-$r$ plane is closest to the horizon at $r_{+}$.
This is because $r(\rho < 0)$ spirals toward $r = r_{+}$ slowly (see, for example, Fig.~\ref{fig:r_from_rho_neg_different_rsmp}).
With a smaller $r_{*}^{\text{mp}}$, the same $\rho$ corresponds to a value of $r = r(\rho)$ that is closer (in the modulus sense) to $r = r_+$. Therefore, the absolute value of the inner numerical boundary $|\rho_{\text{in}}|$ needed when solving the \gls{GSN} equation (cf. Sec.~\ref{subsec:complex_freq_soln}) can be smaller, and, thus, it is more efficient.

In practice, however, one needs to evaluate $e^{\pm ip r_{*}^{\text{mp}}}$ and $e^{\pm i\omega r_{*}^{\text{mp}}}$ during a calculation, which can still be exponentially large or small in modulus. 
Therefore, one should also make sure that their choice of the matching point $r_{*}^{\text{mp}}$ will not make those numbers too big or small.
In the numerical implementation \texttt{GeneralizedSasakiNakamura.jl}, we enforce that $1/25 \leq e^{|\text{Im}\;\omega|r_{*}^{\text{mp}}} \leq 1$ (note that $\text{Im}\;\omega = \text{Im}\;p$).
\\
\bibliography{BHPT}%

\end{document}